\documentclass[rmp,aps,nofootinbib,graphicx]{revtex4-1}

\usepackage{amssymb,amsmath}
\usepackage{graphicx}
\newtheorem{theorem}{Theorem}

\draft 

\begin{document}


\title{Phase transitions and quantum effects in anharmonic crystals  } 



\author{Sergio Albeverio}
\email[]{albeverio@uni-bonn.de}
\affiliation{Institut f\"ur Angewandte Mathematik, Universit\"at Bonn,  53115 Bonn, Germany}

\author{Yuri Kozitsky}
\email[]{jkozi@hektor.umcs.lublin.pl}
\affiliation{Instytut Matematyki, Uniwersytet Marii Curie-Sk{\l}odowskiej, 20-031 Lublin, Poland}
\author{Yuri Kondratiev}
\email[]{kondrat@math.uni-bielefeld.de}
\affiliation{Fakult\"at f\"ur Mathematik, Universit\"at Bielefeld,  33615 Bielefeld, Germany}

\author{Michael R\"ockner}
\email[]{roeckner@math.uni-bielefeld.de}
\affiliation{Fakult\"at f\"ur Mathematik, Universit\"at Bielefeld,  33615 Bielefeld, Germany}


\date{\today}

\begin{abstract}
The most important recent results in the theory of phase transitions and quantum
effects in quantum anharmonic crystals are
presented and discussed. In particular, necessary and sufficient conditions for a
phase transition to occur at some temperature
are given in
the form of simple inequalities involving the interaction strength and the
parameters describing a single oscillator.
The main characteristic feature of the theory is that both mentioned phenomena are
described in one and the same setting, in which
thermodynamic phases of the model appear as probability measures on path spaces.
Then the possibility of a phase transition to occur
is related to the existence of multiple phases at the same values of the relevant
parameters.
Other definitions of phase transitions, based on the non-differentiability of the
free energy density
and on the appearance of ordering, are also discussed.\\[1cm]


\end{abstract}


\maketitle 

\tableofcontents
\section{Introduction and Overview}
\label{Sec1}

Understanding what differentiates the collective behavior of a large quantum system from that of
its classical counterpart is a challenging problem of theoretical physics.
An important example
here is a phase transition, due to which the system becomes ordered in one or another way.
Intuitively, it is
clear that intrinsic randomness of quantum systems should enhance thermal fluctuations
in their work against ordering. For the most of realistic
models, the theoretical explanation of such quantum effects is typically based on additional simplifications
and uncontrolled approximations.
However, as we show in this article, for a certain class of important realistic quantum models
a rigorous mathematical theory based on the so called `first principles' can be developed in detail, qualitatively agreeing with relevant
experimental data. These models are {\it quantum anharmonic crystals}, used, e.g., in the theory of structural phase transitions in solids
triggered by the ordering of light particles. At low temperatures, the mentioned quantum effects prevent such light particles from ordering and thus stabilize the crystal.
We call this effect {\it quantum stabilization}, see \cite{[A1]}.
Its mathematical description is based on the use of infinite dimensional (path) integrals and the corresponding path measures, which serve here as
mathematical models of thermodynamic phases. This allows us to apply powerful tools of modern mathematical analysis, which, however,
makes the theory quite involved.
The purpose of the present article is to explain the main aspects of this theory in the way accessible
also to non-mathematicians. We do this by interpreting the corresponding ideas and results published in our recent monograph \cite{[A3]}.
As we aim at reviewing the results, but not the literature, the quotations are restricted to a necessary minimum.
The reader interested in the bibliography on this topic is cordially invited to find it in the Bibliographic Notes to each chapter of \cite{[A3]}.

An anharmonic oscillator is a mathematical model of a point particle moving in a potential field
with multiple minima and sufficient growth at infinity, which makes the particle motion confined to the vicinity of the point of its (possibly unstable) equilibrium.
In the simplest case, the displacement $q$  of the particle from this point is one-dimensional and the corresponding potential energy $V(q)$
is  a symmetric continuous function with
two minima at $\pm q_*$,
separated by a barrier. In this case, one speaks of {\it double-well} potentials, an example of which is given in Fig. 1.

\begin{figure}[h!]
\centering{\includegraphics[width=0.3\textwidth]{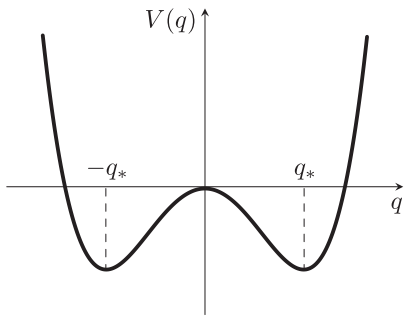}}
\caption{The case of $V(q)=-\alpha q^2+b q^4$, $\alpha$ and $b$ strictly positive.}
\end{figure}

Let now
an infinite system of such point particles be arranged into a crystal. That is, each particle is localized in  the vicinity of
its own crystal site. Suppose also that the particles interact with each other. The corresponding model, called an {\it
anharmonic crystal}, is often used in solid state physics to describe, e.g., ionic crystals containing localized light particles
oscillating in the field created by heavy ionic complexes. The ordering of such light particles may trigger a structural change of the whole crystal, see \cite{[BC]}. A particular example of this sort can be a KDP-type ferroelectric with hydrogen bounds,
in which such light particles are protons or deuterons performing one-dimensional oscillations along the bounds,
see \cite{[BZ],[S],[TM]}. Another relevant physical object is a system of apex oxygen ions in YBaCuO-type
high-temperature superconductors \cite{[Fr],[M],[Stas],[Stas1]}. Anharmonic oscillators are also used in models describing
the interaction of vibrating particles with a radiation field \cite{[Hainzl],[HHS]}, strong electron-electron correlations
caused by the interaction of electrons with vibrating ions \cite{[Freer]}, or charge transfer along hydrogen bounds \cite{[Stas3]}.

If the particle motion obeys the laws of classical mechanics,
the low energy states of the particle are degenerate. By this we mean the
existence of different states with the same energy, as the particle
is confined to one of the wells if its energy is less than the height of the potential barrier
separating the wells.
Then the low temperature
equilibrium thermodynamic states (phases) of the corresponding anharmonic crystal can also be multiple.  If this is the case,
as the temperature changes the crystal undergoes
a phase transition -- a collective phenomenon
caused by the interaction and by the degeneracy mentioned above. If the particle mass is sufficiently small, it
moves according to the laws of quantum mechanics, and hence can tunnel through the potential barrier. This tunneling motion eliminates
the degeneracy, which might
affect the ability of the phase transition to occur or even suppress it completely. On the theoretical level, such `quantum effects' were
first discussed in \cite{[SBS]}.
Later on, a number of
publications on this item appeared, see \cite{[MPZ],[STZ],[VZ1],[VZ]} and the literature quoted therein.
The key conclusion of those works
is that the quantum effects
become strong whenever the particle mass $m_{\rm ph}$ gets small.
This agrees with the experimental data on the isotopic effect in
hydrogen bound ferroelectrics \cite{[BZ]} and in YBaCuO-type high-temperature superconductors \cite{[M]}. At the same time,
experiments show that high hydrostatic pressure applied to a ferroelectric crystal diminishes the phase transition temperature, see, e.g.,
\cite{[Tib]} and
the table on page 11 in \cite{[BZ]}. Hence, the reduction of the distance between the wells amplifies quantum effects.
From this one concludes that
also the shape of the localizing potential $V$
plays an important role here.

In the mathematical theory of classical anharmonic crystals, a thermodynamic phase appears as
a Gibbs measure defined by a specific condition, involving the potential energy of a single particle and the
interaction between the particles. Usually, this condition is formulated as
the {\it Dobrushin-Lanford-Ruelle} {\it (DLR) equation} which the Gibbs measures in question have to solve.
For quantum infinite particle systems, thermodynamic states are usually described as linear functionals
on non-commutative algebras of observables, see \cite{[BR]}. However, for quantum crystals the direct
construction of such states appears to be impossible as the relevant observables are unbounded operators.
In view of this,
the results mentioned above were based on rather indirect indications,
e.g., the lack of a symmetry breaking \cite{[VZ1]} or of an order parameter \cite{[MPZ]}.
It soon had become clear that a mathematical theory, which describes phase transitions
and quantum effects in a unified way, and which as well takes into account the role of
the localizing potential, ought to be based on a precise definition of thermodynamic phases. Such a theory was elaborated in a series
of works, see \cite{[A],[A0],[A00],[A1],[A2],[A4],[K3],[KK],[KKK],[KP],[P]} and the citations therein.
The key aspect of this theory is the use of path integral techniques in which thermodynamic phases are
constructed as probability measures on path spaces (path measures) that solve the corresponding DLR equations.
Thereafter,  necessary and sufficient conditions are derived for a phase transition to occur expressed in terms of
the particle mass $m_{\rm ph}$, the interaction
strength $\hat{J}_0$, and a parameter characterizing the potential energy $V$.
In the simplest case where
\begin{equation}
  \label{00}
V(q)= - \alpha q^2 + b q^4, \quad {\rm and } \ \   q_* = \sqrt{\alpha/2 b},
\end{equation}
cf. Fig.1, both conditions can be expressed through
one and the same parameter $Q=(4m_{\rm ph}/9\hslash^2) q_*^4 \hat{J}_0$, and have the following surprisingly simple form:
(a) $ Q> d \theta(d)$ (sufficient); (b) $ Q\geq \varkappa$ (necessary). Here $\theta(d)$ is a certain function of the lattice
dimension $d$, such that $d\theta (d) >1$ and $d\theta (d) \to 1$ as $d\to +\infty$, see (\ref{q30}) and (\ref{q31}) below.
The parameter $\varkappa\leq 1$ is determined by the spectral properties of the single-particle Hamiltonian (\ref{2}) with $h=0$, see (\ref{S6}) below.

A self-contained presentation of path integral methods in the statistical mechanics
of systems of interacting quantum anharmonic oscillators is given in the monograph \cite{[A3]}, addressed to
both communities --  mathematicians and physicists.  The reader can find here a  collection of facts, concepts, and tools relevant for the application
of functional-analytic and measure-theoretic methods to problems of quantum statistical mechanics. This includes, in particular,
a complete description of the theory mentioned above, as well as of its far-going extensions and implications.
In this article based on that monograph, we present and explain some of its details.

\section{The Model and its Thermodynamics}

\label{SS2}
\subsection{The model}

\label{SS2.1}

To be more concrete, we assume that the considered model describes an ionic crystal and thus adopt
the ferroelectric terminology. This means that we study a system of light particles oscillating in the field created by
ionic complexes, which we suppose to be fixed (adiabatic approximation). A substance modeled in this way might be a KDP type ferroelectric \cite{[BZ],[TM]}, in which such particles
are protons -- ions of hydrogen, constituting hydrogen bounds.
Each particle carries an electric charge; hence, its displacement
from equilibrium produces a dipole moment, proportional to the displacement. Therefore, the main contribution to
the interaction between two particles is proportional to the product of their displacements. According to these arguments,
our model is described by the Hamiltonian
\begin{equation}
  \label{1}
 H = - \frac{1}{2} \sum_{x, y}J_{xy} (q_x, q_y) + \sum_{x} H_x .
\end{equation}
Here the sums run through a crystalline lattice, which we assume to be a $d$-dimensional
`hypercubic' lattice $\mathbb{Z}^d$ consisting of vectors $x= (x_1, \dots, x_d)$, $x_i\in \mathbb{Z}$, and $\mathbb{Z}$ being the set of all (positive and non-positive) integers. The displacement $q_x$ of the particle localized near such
$x= (x_1 , \dots , x_d)$ is supposed to be
a $\nu$-dimensional vector, i.e., $q_x = (q_x^{(1)}, \dots , q_x^{(\nu)})$ and
\begin{equation}
  \label{sp}
(q_x, q_y) = \sum_{j=1}^\nu q_x^{(j)} q_y^{(j)}
\end{equation}
stands for the corresponding scalar product.
Regarding the interaction intensities $J_{xy}$, we suppose that there exists a real-valued function $\varphi$ such that
\begin{equation}
  \label{1a}
 J_{xy} = \varphi (|x_1-y_1|, \dots, |x_d- y_d|).
\end{equation}
We say that the interaction has {\it finite range} if $\varphi$ vanishes whenever either of
its arguments exceeds a certain value. In particular, $\varphi$ can be nonzero only if
one of its arguments is equal to one and all other arguments are equal to zero. This is the case of a {\it nearest neighbor interaction}.
To avoid unnecessary mathematical complications, in this article we suppose that $J_{xy}$ has finite range. Note, however, that in \cite{[A3]} we consider the general case
with the only assumption that
\begin{equation}
  \label{J}
  \hat{J}_0 := \sum_{y\in \mathbb{Z}^d} |J_{xy}| < \infty,
\end{equation}
see also \cite{[KP]}.
In addition, we suppose that $J_{xy} \geq 0$, which means that
we consider ferroelectric interactions only. The single-particle Hamiltonian in (\ref{1}) has the following form
\begin{eqnarray}
  \label{2}
H_x & = & \frac{1}{2m} |p_x|^2 + V(q_x) - (h,q_x)\\[.2cm]
& = & H_{x}^{\rm har} + U(q_x) - (h,q_x), \nonumber
\end{eqnarray}
where $h\in \mathbb{R}^\nu$ is an external (electric) field,
\begin{equation}
\label{200}
 H_{x}^{\rm har} =  \frac{1}{2m} |p_x|^2 + \frac{a}{2} |q_x|^2,
\end{equation}
is the Hamiltonian of a harmonic oscillator of rigidity $a>0$, and
\begin{equation}
  \label{2001}
U(q_x) = V(q_x) - \frac{a}{2} |q_x|^2,
\end{equation}
is the anharmonic part of the potential energy $V$. For the sake of convenience, we exclude from the latter
the external field term.
One observes that $H_x$  as given in (\ref{2}) is independent of the choice of $x$;
thus, the model (\ref{1}) is invariant under the translations of the lattice $\mathbb{Z}^d$.
The mass parameter $m$ in (\ref{2}) and (\ref{200}) includes Planck's constant, i.e.,
\begin{equation}
  \label{2a}
  m= m_{\rm ph}/\hslash^2.
\end{equation}
The Hamiltonian (\ref{2}) is defined as a linear operator acting in the `physical' complex Hilbert space $\mathcal{H}_x = L^2(\mathbb{R}^\nu)$ of square-integrable `wave' functions,
which have the standard physical interpretation. In particular, for $\psi \in \mathcal{H}_x$ such that
\begin{equation*}
\|\psi \|^2_{\mathcal{H}_x} := \int_{\mathbb{R}^\nu} |\psi(u)|^2 du =1,
\end{equation*}
and for a bounded linear operator $A: \mathcal{H}_x \to \mathcal{H}_x$, the (complex) number
\begin{equation}
  \label{2aa}
\omega_\psi (A) := (\psi , A \psi)_{\mathcal{H}_x} = \int_{\mathbb{R}^\nu} \bar{\psi}(u) (A\psi)(u)du
\end{equation}
is said to be the value of $A$ in vector state $\omega_\psi$. Here
\[
(\phi, \psi)_{\mathcal{H}_x} = \int_{\mathbb{R}^\nu} \bar{\phi}(u) \psi(u) du, \quad \ \ \phi , \psi \in \mathcal{H}_x,
\]
is the scalar product
in $\mathcal{H}_x$, and $\bar{\phi}$ stands for the complex conjugate.
In general, a {\it state}, $\omega$, is a map defined  on the algebra $\mathfrak{C}_x$ of bounded linear operators
$A: \mathcal{H}_x \to \mathcal{H}_x$ with values in the set of complex numbers $\mathbb{C}$  such that, for all
$\varsigma_1, \varsigma_2 \in \mathbb{C}$ and all operators  $A_1 , A_2 \in \mathfrak{C}_x$,
\begin{eqnarray}
\label{ST}
& & \omega(\varsigma_1 A_1 + \varsigma_2 A_2) = \varsigma_1 \omega (A_1) + \varsigma_2 \omega (A_2), \\[.2cm]
& & \ \ \omega(I) =1, \quad \  \omega (A) \geq 0 \ \ \ {\rm for} \ \ \ A\geq 0. \nonumber
\end{eqnarray}
Here $I$ is the identity operator and $A\geq 0$ means that $(\psi , A \psi)_{\mathcal{H}_x}\geq 0$ for all $\psi \in\mathcal{H}_x$.
Let $\omega_1$ and $\omega_2$ be two states. For $\varsigma\in [0,1]$, we set
\begin{equation}
  \label{01}
\omega = \varsigma \omega_1 + (1-\varsigma) \omega_2.
\end{equation}
Then $\omega$ is also a states, which is a convex combination ({\it mixture}) of $\omega_1$ and $\omega_2$.
The combination in (\ref{01}) is called nontrivial, if $\omega \neq \omega_1$ and $\omega \neq \omega_2$.
By definition, a pure state is a state which cannot be decomposed into a nontrivial convex combination of other states.
Vector states (\ref{2aa}) are pure, see Theorem 1.1.15, page 26 in \cite{[A3]}.

In the sequel, we mostly consider the following versions of the model (\ref{1}), (\ref{2}):
\vskip.1cm
\begin{tabular}{ll}
(i) \ &the dimension $\nu$ is arbitrary and $V$ has the form\\[.1cm]
&cf. (\ref{00}),
\end{tabular}
\begin{equation}
  \label{3}
 V(q) = - \alpha |q|^2 + b |q|^4, \qquad \alpha , b >0;
\end{equation}
\vskip.1cm
\begin{tabular}{ll}
(ii) \ &$\nu=1$ and $V$ is an even continuous function\\[.1cm] &such that, for $|q| < |\tilde{q}|$,
\end{tabular}
\vskip.1cm
\begin{equation}
  \label{3a}
W(q) - V(q) \leq W(\tilde{q}) - V(\tilde{q}),
\end{equation}
\begin{tabular}{ll}
\hskip1.4cm
&\qquad \ where
\end{tabular}
\begin{eqnarray}
  \label{3b}
 & & W(q) = - b_1q^2+ \sum_{l=2}^r b_l q^{2l}, \\[.2cm]& & b_1, b_r > 0, \ \ b_l \geq 0, \ {\rm for} \ l=2, \dots , r-1, \nonumber
\end{eqnarray}
\vskip.1cm
\begin{tabular}{ll}
 &and $r\geq 2$ is a finite integer;\\[.2cm]
(iii) \ &$\nu=1$ and $V$ is continuous, asymmetric, i.e.,\\[.1cm] &$V(q) \neq V(-q)$, and such that the following\\[.1cm] &estimate holds
\end{tabular}
\begin{equation}
  \label{3ba}
V(q) \geq - C + D|q|^{2 + \epsilon},  \quad  q\in \mathbb{R},
\end{equation}
\vskip.1cm
\begin{tabular}{ll}
&\qquad where $C$, $D$, and $\epsilon$ are positive constants.
\end{tabular}
\vskip.1cm \noindent
Assuming that $V$ in (\ref{3a}) is differentiable, we can rewrite this condition in the form
\begin{equation*}
 - b_1 + \sum_{l=2}^r l b_l \vartheta^{l-1} \geq v'(\vartheta), \qquad {\rm for} \ {\rm all} \ \ \vartheta >0,
\end{equation*}
where $v$ is such that $V(q)= v(q^2)$.
Along with the function $W$ we also use
\begin{equation}
  \label{3c}
  \Phi(\vartheta) = \sum_{l=2}^r \frac{(2l)!}{2^{l-1}(l-1)!}b_l \vartheta^l,\quad \vartheta \geq 0,
\end{equation}
where $r$ and  all $b_l$ are the same as in (\ref{3b}).
One observes that in cases (i) and (ii) with $h=0$, the model is $O(\nu)$-symmetric, that is, it is symmetric with respect to all orthogonal transformations of $\mathbb{R}^\nu$,
which constitute the group\footnote{Note that the only nontrivial transformation in $O(1)$ is $q \mapsto -q$.} $O(\nu)$. Then the phase transition in the corresponding model
can be associated with the breakdown of this symmetry. Case (iii) is essentially
different from this point of view.

We conclude this subsection by recalling that the Hamiltonian in  (\ref{1}) corresponds to a quantum model. In particular, this means
that $p_x$ and $q_x$ are (unbounded) operators, which in view of (\ref{2a})  obey the following commutation relations
\begin{equation}
  \label{3dd}
p^{(j)}_x q^{(j')}_y - q^{(j')}_y p^{(j)}_x = - {\rm i} \delta_{xy}\delta_{jj'}, \qquad {\rm i} = \sqrt{-1}.
\end{equation}

\subsection{A single anharmonic oscillator}

\label{SS2.2}

As we shall see below, important thermodynamic properties of the model (\ref{1}) are closely related
to the quantum-mechanical properties of the  anharmonic oscillator described by the Hamiltonian (\ref{2}), in which we set $h=0$.
To avoid nonessential technical problems,
 we suppose that $V$ in (\ref{2}) is an $O(\nu)$-symmetric polynomial and thus is
of the following form
\begin{equation}
\label{V1}
V(q) = \sum_{l=1}^r c_l |q|^{2l}, \qquad c_r>0,
  \end{equation}
with $r\geq 1$, cf. (\ref{3}). This, of course, includes also the case of the harmonic oscillator (\ref{200}).
For such $V$, there exists a subset $\mathcal{D}$ of the physical Hilbert space
$\mathcal{H}_x$
such that $H_x$ with domain $\mathcal{D}$ is a self-adjoint operator, see Theorem 1.1.47, page 46 in \cite{[A3]}.
Its spectrum consists entirely of real eigenvalues $E_n$, $n\in \mathbb{N}_0$, such that, for some $E_*>0$ and $n_*\in \mathbb{N}$,
they obey the estimate
\begin{equation*}
  E_n \geq E_* n^{2r/(r+1)}, \qquad {\rm for} \ \ {\rm all} \ \ n\geq n_*,
\end{equation*}
see Theorem 1.1.58, page 53 in \cite{[A3]}. In particular, the eigenvalues $E^{\rm har}_n$, $n\in \mathbb{N}_0$, of $H_x^{\rm har}$, are
\begin{equation}
  \label{V3}
E_n^{\rm har} = (n+ \nu/2)\Delta^{\rm har}, \qquad \Delta^{\rm har} := \sqrt{a/m},
\end{equation}
see e.g. Proposition 1.1.37, page 41 in \cite{[A3]}. If $\nu=1$, the eigenvalues $E_n$
are simple, which means that, for every $n\in \mathbb{N}_0$, the equation
\[
H_x \psi_n = E_n\psi_n
\]
has exactly one solution $\psi_n \in \mathcal{H}_x$.
The dynamical properties of the oscillator crucially depend on the following {\it gap} parameter
\begin{equation}
  \label{V5}
 \Delta = \inf_{n\in \mathbb{N}_0} \left( E_{n+1}- E_n\right).
\end{equation}
Note that
\begin{equation}
  \label{V55}
 \Delta \to 0 , \qquad \ {\rm as} \ \ m \to +\infty.
\end{equation}
The latter, in fact, is the classical limit $\hslash \to 0$, see (\ref{2a}).
It can be shown that, for a fixed value of the mass parameter $m$,
\begin{equation}
  \label{V4}
 \lim_{n\to +\infty} \left( E_{n+1}- E_n\right) = + \infty.
\end{equation}
Thus, the infimum in (\ref{V5}) is strictly positive for finite values of $m$.
As usual, for appropriate functions $f$ and $g$, by writing $f \sim g$ we mean that $\lim (f/g) = 1$
assuming that the meaning of the limit is clear from the context.
In the harmonic case, the gap parameter is given in (\ref{V3}).
According to Theorem 1.1.60, page 59 in \cite{[A3]},
$\Delta$ is a continuous function of the mass parameter $m>0$ such that, cf. (\ref{V55}),
\begin{equation}
\label{V6}
\Delta \sim \Delta_0 m^{-r/(r+1)}, \qquad \ \  {\rm as} \ \ m\to 0,
\end{equation}
where $r$ is the same as in (\ref{V1}) and (\ref{3b}).
The proof of these properties goes as follows. First, by the analytic perturbation theory for
self-adjoint operators one shows that the eigenvalues $E_n$, and hence the increments $E_{n+1}- E_n$, continuously depend on $m$.
This, however, does not yet mean the continuity of $\Delta$, as the infimum of an infinite number of continuous functions need not be continuous.
But,
by (\ref{V4}) we see that, for some  $n_0\in \mathbb{N}$,
\[
E_{n+1} - E_n > E_{k+1} - E_k
\]
for any $n> n_0$ and any $k< n_0$. This means that the infimum in (\ref{V5}) can be found by means of a finite number of increments, i.e.,
$$\Delta = \min_{n= 0, 1, \dots , n_0}\left( E_{n+1}- E_n\right),$$ which yields the continuity in question.
Finally, the asymptotics (\ref{V6}) is obtained by means of a rescaling and then by the asymptotic perturbation theory for self-adjoint operators. From this we conclude that the following parameter
\begin{equation}
  \label{V7}
R_m := m \Delta^2
\end{equation}
is a continuous function of $m$ and
\begin{equation}
  \label{V8}
  R_m \sim \Delta_0^2 m^{-(r-1)/(r+1)}, \qquad \ \  {\rm as} \ \ m\to 0.
\end{equation}
Noteworthy, $R_m$, $\hat{J}_0$ as given in (\ref{J}), and the rigidity $a$ as in (\ref{200}), are measured in the same physical units, cf. (\ref{2a}). We also note that $R_m \to 0$ in the classical limit $m\to +\infty$, see (\ref{S5}) below.
In the harmonic case, $R_m^{\rm har} =a$, see (\ref{V3}). By analogy, we call $R_m$ the {\it quantum rigidity} of the oscillator described by $H_x$. By the mentioned continuity and by (\ref{V8}), for $r\geq 2$,
$R_m$ takes any value from an interval $[R^{(0)}, +\infty)$ for some $R^{(0)}>0$. As we shall see below,
for big enough values of $R_m$, phase transitions in the model (\ref{1}) are suppressed at all temperatures.

\subsection{Thermodynamic phases}
\label{SS2.3}

The Hamiltonian given in (\ref{1}) has no direct mathematical meaning and serves as a formula for `local' Hamiltonians
\begin{equation}
  \label{3e}
  H_\Lambda = - \frac{1}{2} \sum_{x, y\in \Lambda}J_{xy} (q_x, q_y) + \sum_{x\in \Lambda} H_x,
\end{equation}
indexed by finite subsets $\Lambda \subset \mathbb{Z}^d$. Let $|\Lambda|$ stand for the number of lattice sites contained in $\Lambda$.
Then the corresponding
physical Hilbert space
is $\mathcal{H}_\Lambda = L^2 ((\mathbb{R}^{\nu})^{|\Lambda|})$. Among all linear operators acting in $\mathcal{H}_\Lambda$, we distinguish
bounded operators $A:\mathcal{H}_\Lambda \to \mathcal{H}_\Lambda$. The set of all such operators $\mathfrak{C}_\Lambda$ is
a $C^*$-algebra, see Sections 1.1 and 1.2 in \cite{[A3]} for more information on this topic.
The local Hamiltonian defined in (\ref{3e})
is an unbounded operator. However,
by the assumptions made above $H_\Lambda$ is a self-adjoint operator on some domain $\mathcal{D}_\Lambda \subset \mathcal{H}_\Lambda$. Moreover, one can show that, for any $\beta >0$,
the operator $\exp\left( - \beta H_\Lambda\right)$ has finite trace, and hence the local partition function
\begin{equation}
  \label{4}
 Z_\Lambda := {\rm trace} \exp\left( - \beta H_\Lambda\right) ,
\end{equation}
is well-defined.
Along with the Hamiltonian (\ref{3e}) we will use `periodic' Hamiltonians defined as follows.
Let $\Lambda_L$, $L\in \mathbb{N}$, be the subset of the lattice $\mathbb{Z}^d$ consisting of all $x=(x_1, \dots , x_d)$
such that $-L < x_i \leq L$, for all $i=1, \dots, d$. For $x_i$ and $y_i$,  obeying the latter condition, we set
\begin{equation*}
|x_i - y_i|_L  =  \min\{|x_i - y_i|; 2L - |x_i - y_i| \}.
\end{equation*}
Clearly, $\Lambda_L$ is a cube, which one can turn into a torus by identifying
its opposite walls. We denote it also by $\Lambda_L$ and equip with
the distance
\begin{equation}
  \label{4z}
|x-y|_{L}  =   \sqrt{ |x_1 - y_1|_L^2 + \cdots + |x_d - y_d|_L^2}.
\end{equation}
After that, $\Lambda_L$ becomes invariant with respect to the corresponding translations.
Now we introduce
\begin{equation*}
J^L_{xy} = \varphi(|x_1- y_1|_L , \dots , |x_d- y_d|_L) , \quad x, y \in \Lambda_L,
\end{equation*}
where $\varphi$ is as in (\ref{1a}). Then the periodic Hamiltonian in $\Lambda_L$ is
\begin{equation}
  \label{4b}
H_L =  - \frac{1}{2} \sum_{x, y\in \Lambda_L}J^L_{xy} (q_x, q_y) + \sum_{x\in \Lambda_L} H_x.
\end{equation}
Like the Hamiltonian in (\ref{3e}), it is a  self-adjoint operator in the corresponding space $\mathcal{H}_{\Lambda_L}$,
for which
\begin{equation}
  \label{4c}
 Z_L := {\rm trace} \exp\left( - \beta H_L \right) < \infty.
\end{equation}
The local Gibbs state $\varrho_\Lambda$ at temperature $T = 1/ \beta k_B$, $k_B$ being Boltzmann's constant, is defined to be
\begin{equation}
  \label{5}
\varrho_\Lambda (A) =   {\rm trace}\left\{ A \exp\left( - \beta H_\Lambda\right)\right\}/Z_\Lambda, \quad A \in \mathfrak{C}_\Lambda.
\end{equation}
It is a positive linear functional on the algebra $\mathfrak{C}_\Lambda$, cf. (\ref{ST}).
Moreover, it is the following mixture, cf. (\ref{01}),
\[
\varrho_\Lambda = \frac{1}{Z_\Lambda}\sum_{n=0}^\infty \exp\left( - \beta E^{(n)}_\Lambda\right) \omega_{\psi_n}
\]
of the vector states $\omega_{\psi_n}$ corresponding to the eigenvectors of $H_\Lambda$ defined by
\[
H_\Lambda \psi_n = E^{(n)}_\Lambda \psi_n.
\]
In the same way, we define
the periodic local state
\begin{equation}
  \label{5a}
  \varrho_L (A) =   {\rm trace}\left\{ A \exp\left( - \beta H_L \right)\right\}/Z_L, \quad A \in \mathfrak{C}_\Lambda.
\end{equation}
If not explicitly stated otherwise, all properties of the states $\varrho_\Lambda $ that we mention
in the sequel are attributed also to the periodic states $\varrho_L$.

By H{\o}egh-Krohn's theorem, see page 72 in \cite{[A3]}, $\varrho_\Lambda$ can be recovered from its values on the
 products of the form
\begin{eqnarray*}
& & \qquad \mathfrak{a}^\Lambda_{t_1} (F_1) \cdots \mathfrak{a}^\Lambda_{t_n} (F_n), \quad n\in \mathbb{N}, \\[.2cm]  & &  F_1 , \dots , F_n \in \mathfrak{F}_\Lambda,\ \qquad
t_1 , \dots , t_n \in \mathbb{R}, \nonumber
\end{eqnarray*}
where $\mathfrak{F}_\Lambda$ is a `rich enough' family of multiplication operators by bounded measurable functions $F:\mathbb{R}^{|\Lambda|} \to \mathbb{C}$. Here
\begin{equation*}
 \mathfrak{a}^\Lambda_{t} (A) = \exp( {\rm i}t H_\Lambda) A \exp( -{\rm i}t H_\Lambda), \qquad  A\in \mathfrak{C}_\Lambda,
\end{equation*}
is the time automorphism $\mathfrak{a}^\Lambda_{t} : \mathfrak{C}_\Lambda \to \mathfrak{C}_\Lambda$ which in the Heisenberg approach describes the dynamics of the oscillators located in $\Lambda$.
Thus, the Green functions
\begin{equation}
  \label{7}
G_{F_1, \dots , F_n} (t_1 , \dots , t_n) := \varrho_\Lambda \left[\mathfrak{a}^\Lambda_{t_1} (F_1) \cdots \mathfrak{a}^\Lambda_{t_n} (F_n) \right],
\end{equation}
with all possible choices of $n\in \mathbb{N}$ and then of $F_1 , \dots , F_n \in \mathfrak{F}_\Lambda$ determine the state $\varrho_\Lambda$.
Each Green function admits an analytic continuation,
also denoted by $G_{F_1, \dots , F_n} $,
to the domain ${D}_n$ consisting of those $(\zeta_1 , \dots , \zeta_n) \in \mathbb{C}^n$, for which
\[
0 < {\rm Im}(\zeta_1 ) < \cdots < {\rm Im}(\zeta_n ) < \beta.
\]
Furthermore, see Theorem 1.2.32, page 78 in \cite{[A3]}, $G_{F_1, \dots , F_n} $
is continuous on the closure $\overline{{D}}_n$ and can uniquely be recovered from its restriction to
the set
\[{D}^{(0)}_n := \{ (\zeta_1 , \dots , \zeta_n) \in \overline{\mathcal{D}}_n :
 {\rm Re}(\zeta_1 ) = \cdots = {\rm Re}(\zeta_n )=0\}.
\]
Thus, the Matsubara functions
\begin{equation}
  \label{8}
 \Gamma_{F_1, \dots , F_n} (\tau_1 , \dots , \tau_n) := G_{F_1, \dots , F_n} ({\rm i}\tau_1 , \dots ,{\rm i}\tau_n),
\end{equation}
with
\begin{equation}
\label{8y}
0\leq \tau_1 \leq \tau_2 \cdots \leq \tau_n \leq \beta,
\end{equation}
and with all possible choices of $n\in \mathbb{N}$ and then of $F_1 , \dots , F_n \in \mathfrak{F}_\Lambda$, uniquely determine the state (\ref{5}). By (\ref{7}), we can rewrite (\ref{8}) in the form
\begin{widetext}
\begin{equation}
\label{8z}
\Gamma_{F_1, \dots , F_n} (\tau_1 , \dots , \tau_n) =
{\rm trace}\left[F_1 e^{-(\tau_2 - \tau_1)H_\Lambda}F_2 e^{-(\tau_3 - \tau_2)H_\Lambda} \cdots F_n e^{-(\tau_{n+1} - \tau_n)H_\Lambda}\right]  /Z_\Lambda,
\end{equation}
\end{widetext}
where $\tau_{n+1} = \beta + \tau_1$ and  the arguments obey (\ref{8y}).
The main ingredient of our technique is the following representation, see Theorem 1.4.5 in \cite{[A3]},
\begin{eqnarray}
  \label{9}
& & \Gamma_{F_1, \dots , F_n} (\tau_1 , \dots , \tau_n)\\[.2cm] & &   = \int_{\Omega_\Lambda} F_1 (\omega_\Lambda (\tau_1)) \cdots F_n (\omega_\Lambda (\tau_n)) \nu_\Lambda (d \omega_\Lambda), \nonumber
\end{eqnarray}
where $\nu_\Lambda$ is the `local Euclidean Gibbs measure' in $\Lambda$, which is a probability measure on the measurable space
$(\Omega_\Lambda , \mathcal{B}(\Omega_\Lambda))$. Here
\begin{eqnarray*}
 & & \qquad \Omega_\Lambda  :=  \{\omega_\Lambda = (\omega_x)_{x\in \Lambda}: \omega_x \in C_\beta \}, \\[.2cm]
 & & C_\beta  :=  \{ \phi\in C([0,\beta]\to \mathbb{R}^\nu): \phi (0) = \phi(\beta)\},
\end{eqnarray*}
and $C([0,\beta]\to \mathbb{R}^\nu)$ stands for the set of all continuous functions (paths) $\phi:[0,\beta] \to \mathbb{R}^\nu$, also called
`continuous
temperature loops' in view of the periodicity property $\phi(0 ) = \phi (\beta)$.
Thus, $ C_\beta$ is the Banach space of paths. The space  $\Omega_\Lambda$ is equipped with the corresponding product topology and with the Borel $\sigma$-field
$\mathcal{B}(\Omega_\Lambda)$. The measure $\nu_\Lambda$ has the following Feynman-Kac representation
\begin{equation}
  \label{10}
 \nu_\Lambda (d\omega_\Lambda) = \exp\left[ - I_\Lambda (\omega_\Lambda)\right]\chi_\Lambda (d \omega_\Lambda) /N_\Lambda(h),
\end{equation}
in which
\begin{eqnarray}
  \label{11}
  I_\Lambda (\omega_\Lambda) & = & - \frac{1}{2}\sum_{ x, y \in \Lambda}J_{xy} \int_0^\beta (\omega_x (\tau) ,\omega_y (\tau)) d\tau
\\[.2cm] & - & \sum_{x\in \Lambda} \int_0^\beta (h, \omega_x(\tau)) d \tau    +  \sum_{x\in \Lambda}
  \int_0^\beta U(\omega_x (\tau))d \tau \nonumber
\end{eqnarray}
is the energy functional. We recall that $U$ in (\ref{11}) is the anharmonic part of $V$, see (\ref{2001}), and that, for each $\tau$ and $x$, $\omega_x(\tau)$ is a $\nu$-dimensional vector and thus $(\cdot, \cdot)$ in (\ref{11}) stands for the corresponding scalar product, cf. (\ref{sp}). The normalizing term $N_\Lambda(h)$ is defined by the condition
\begin{equation}
  \label{11z}
N_\Lambda (h) =   \int_{\Omega_\Lambda} \exp\left[- I_\Lambda (\omega_\Lambda) \right]\chi_\Lambda (d \omega_\Lambda),
\end{equation}
and the measure $\chi_\Lambda$ has the form
\begin{equation}
  \label{12}
\chi_\Lambda (d\omega_\Lambda) = \prod_{x \in \Lambda} \chi(d\omega_x).
\end{equation}
Here $\chi$ is the H{\o}egh-Krohn measure -- a Gaussian probability measure on the Banach space $C_\beta$, constructed by means
of the harmonic part of
the Hamiltonian (\ref{2}). It is defined by its Fourier transform
\begin{widetext}
  \begin{equation*}
 \int_{C_\beta} \exp\left({\rm i} \sum_{j=1}^\nu \int_0^\beta \lambda^{(j)}(\tau) \omega_x^{(j)}(\tau)d \tau \right)\chi(d \omega_x) = \exp\bigg{(} - \frac{1}{2} \sum_{j=1}^\nu \int_0^\beta \int_0^\beta S(\tau, \tau') \lambda^{(j)}(\tau) \lambda^{(j)}(\tau') d\tau d\tau' \bigg{)},
  \end{equation*}
\end{widetext}
where $\lambda$ is in $C_\beta$ and $S(\tau, \tau')$ is the {\it propagator} corresponding to a single scalar ($\nu=1$) harmonic oscillator,
which, in fact, is the corresponding Matsubara function (\ref{8}) for $n=2$ and $F_1 = F_2 = q_x$. That is, cf. (\ref{8z}), (\ref{200}), and (\ref{V3}),
\begin{widetext}
  \begin{eqnarray}
    \label{204}
 S(\tau, \tau')& = & \frac{1}{Z^{\rm har}}{\rm trace} \bigg{[}q_x \exp\left( - |\tau' - \tau|H^{\rm har}_x \right)q_x \exp\left( - (\beta -|\tau' - \tau|)H^{\rm har}_x \right) \bigg{]} \\[.3cm]
 & = & \bigg{[} \exp\left( - |\tau' - \tau|\Delta^{\rm har} \right) + \exp\left( - (\beta -|\tau' - \tau|)\Delta^{\rm har}\right)  \bigg{]}\bigg{/}
  2 \sqrt{ma} \left[ 1 - \exp\left( - \beta\Delta^{\rm har}  \right) \right],\nonumber
  \end{eqnarray}
\end{widetext}
 see pages 99 and 125 in \cite{[A3]}. In a similar way, one defines also periodic local
Gibbs measures, cf. (\ref{4b}),
\begin{equation}
  \label{10a}
\nu_L (d\omega_{\Lambda_L}) = \exp\left[ - I_L (\omega_{\Lambda_L})\right]\chi_{\Lambda_L} (d \omega_{\Lambda_L}) /N_L(h),
\end{equation}
with
\begin{equation}
  \label{11b}
  N_L(h) = \int_{\Omega_{\Lambda_L}} \exp\left[ - I_L (\omega_{\Lambda_L})\right]\chi_{\Lambda_L} (d \omega_{\Lambda_L}),
\end{equation}
and
\begin{eqnarray*}
  I_L (\omega_{\Lambda_L})& = & - \frac{1}{2} \sum_{  x, y \in {\Lambda_L}}J^L_{xy} \int_0^\beta (\omega_x (\tau) ,\omega_y (\tau)) d\tau
   \\[.2cm]& - & \sum_{x\in \Lambda_L} \int_0^\beta (h, \omega_x(\tau)) d \tau  +  \sum_{x\in {\Lambda_L}}
  \int_0^\beta U(\omega_x (\tau))d \tau. \nonumber
\end{eqnarray*}
Thus, the representation (\ref{9}) leads to
the description of the states (\ref{5}), (\ref{5a}) in terms of
Gibbs measures, similarly as in the case of classical anharmonic crystals. Here, however, each $\nu$-dimensional vector
$q_x$ is replaced by a continuous $\nu$-dimensional path $\omega_x$, which is an element of an infinite dimensional vector space.
Going further in this direction, one  can define  `global' Gibbs states of the model (\ref{1}) as the probability measures
on the space of `tempered configurations' satisfying the DLR equation, see Chapter 3 in \cite{[A3]}.
It can be shown that the set of all such measures, which we denote by $\mathcal{G}$, is a nonempty compact simplex with a nonempty extreme boundary ${\rm ex}( \mathcal{G})$,
the elements of which correspond to the {\it thermodynamic phases} of our model, see also Chapter 7 in \cite{[Ge]}. Thus, by
a thermodynamic phase we understand an extreme element of $\mathcal{G}$. The latter means that such a measure cannot be decomposed
into a nontrivial convex combination of other elements of $\mathcal{G}$. By virtue of the DLR equation, the set $\mathcal{G}$
can contain either one or infinitely many elements.
Correspondingly, the multiplicity (resp. the uniqueness) of thermodynamic phases
existing at a given value of the temperature means that $|{\rm ex}( \mathcal{G})|>1$ (resp. $|\mathcal{G}|=1$).
In general, there exist several ways of establishing whether the set of Gibbs measures is infinite or containing a single element. One of them consists in showing that there exists
an element of $\mathcal{G}$, which is `less symmetric' than the corresponding Hamiltonian, i.e., that a symmetry breaking occurs.
Another way is to show that $\mathcal{G}$ contains an element, which is not in ${\rm ex}(\mathcal{G})$. Since the latter is a subset of $\mathcal{G}$,
then $\mathcal{G}$ is not a singleton, and hence $|\mathcal{G}|=+\infty$.
The latter way will be discussed below in more detail.

\subsection{The free energy}

\label{SS2.4}

A more `traditional'  way of establishing phase transitions is based on the use of thermodynamic functions. In our case, this will be the free energy.
Recall that we assume $J_{xy}$ to be of finite range.
Along with the energy functional (\ref{11}) we consider also
\begin{eqnarray}
  \label{11a}
I_\Lambda (\omega_\Lambda|\xi) & = &  I_\Lambda (\omega_\Lambda)\\[.2cm]& - &
\sum_{x\in \Lambda, \ y \in \Lambda^c}J_{xy}\int_0^\beta (\omega_x (\tau), \xi_y(\tau) )d \tau , \nonumber
\end{eqnarray}
where $I_\Lambda (\omega_\Lambda)$ is as in (\ref{11}) and all $\xi_y$ with $y\in \Lambda^c := \mathbb{Z}^d \setminus \Lambda$, belong to $C_\beta$.
The last term in (\ref{11a}) describes the interaction of the particles in $\Lambda$ with some `environment' defined by the configuration $\xi$ fixed outside $\Lambda$.
We recall that $I_\Lambda (\omega_\Lambda)$ depends on the external field parameter $h\in \mathbb{R}^d$. For $\chi_\Lambda$ as in (\ref{12}), we set \begin{equation*}
N_\Lambda (h, \xi) = \int_{\Omega_\Lambda} \exp\left[-I_\Lambda (\omega_\Lambda|\xi) \right]\chi_\Lambda (d \omega_\Lambda),
\end{equation*}
which is well-defined as $J_{xy}$ has finite range, and hence the sum in (\ref{11a}) is finite.
Obviously, if $\xi_y \equiv 0$ for all $y\in \Lambda^c$, then the above $N_\Lambda(h,\xi)$ coincides with $N_\Lambda(h)$ given in (\ref{11z}).
We also note that the partition function (\ref{4}) and $N_\Lambda(h)$ are related to each other by
\begin{equation}
  \label{q2}
 Z_\Lambda = N_\Lambda (h) Z_\Lambda^{\rm har},
\end{equation}
where $Z_\Lambda^{\rm har}$ is the partition function of the system of non-interacting harmonic oscillators located in $\Lambda$. Explicit
calculations yield, see page 139 in \cite{[A3]},
\begin{eqnarray*}
Z_\Lambda^{\rm har} & = & \left[{\rm trace}\exp\left( - \beta H_x^{\rm har}\right) \right]^{|\Lambda|}\\[.2cm] &=&
\left[ \frac{e^{-\beta \varDelta^{\rm har}/2}}{1 - e^{-\beta \varDelta^{\rm har}}}\right]^{\nu |\Lambda|},  \nonumber
\end{eqnarray*}
where $\Delta^{\rm har} = \sqrt{a/m}$ is the same as in (\ref{V3}) and (\ref{204}). In a similar way, we have
\begin{equation}
  \label{q4}
Z_L = N_L(h)   Z_\Lambda^{\rm har},
\end{equation}
where $Z_L$ and $N_L(h)$ are as in (\ref{4c}) and (\ref{11b}), respectively. Therefore, $N_L(h)$, $N_\Lambda (h)$, and
$N_\Lambda (h,\xi)$ are relative partition functions, and the latter one corresponds to the system of particles in $\Lambda$
interacting with the `environmental' configuration $\xi$. Let us now define
\begin{eqnarray}
  \label{q5}
F_\Lambda (h, \xi) & = & - \frac{1}{\beta |\Lambda|}\ln N_\Lambda (h,\xi), \\[.2cm]
F_\Lambda (h) & = & - \frac{1}{\beta |\Lambda|}\ln N_\Lambda (h), \nonumber
\end{eqnarray}
and
\begin{equation}
  \label{q6}
  F_L (h) = - \frac{1}{\beta |\Lambda_L|}\ln N_L (h).
\end{equation}
Note that $F_\Lambda (h)= F_\Lambda (h, 0)$ and $|\Lambda_L| = (2L)^d$. The functions just introduced are called {\it local free
energy densities}. For a Gibbs measure $\mu\in \mathcal{G}$, we then set
\begin{equation}
  \label{q7}
F^\mu_\Lambda(h) = \int_{\Omega} F_\Lambda (h, \xi) \mu(d\xi).
\end{equation}
This is the free energy density of the system of particles in $\Lambda$, interacting with the environment averaged with respect to $\mu$.
In the following, we are interested in the thermodynamic limits of the functions (\ref{q5}) -- (\ref{q7}). To define such limits, we need
one more notion which we introduce now. Let $\Lambda$ be a finite subset of $\mathbb{Z}^d$. We say that $y\in \mathbb{Z}^d$ is a {\it neighbor}
of $\Lambda$ if: (a) $y$ is not in $\Lambda$, i.e., $y\in \Lambda^c$; (b) there exists $x\in \Lambda$ such that $x$ and $y$ are neighbors, i.e., $$|x-y| = \sqrt{|x_1-y_1|^2 +\cdots + |x_d-y_d|^2} =1.$$ By $\partial \Lambda$ we denote the set of all neighbors of $\Lambda$,
$|\partial \Lambda|$ standing for their number. A sequence $\{\Lambda_n\}_{n\in \mathbb{N}}$ of finite subsets of
$\mathbb{Z}^d$ is said to be a {\it van Hove sequence}, cf. page 193 in \cite{[A3]}, if: (a) $\Lambda_{n} \subset \Lambda_{n+1}$ for every $n\in \mathbb{N}$; (b)
for every $y\in \mathbb{Z}^d$, one finds $n$ such that $y\in \Lambda_n$; (c) $\lim_{n\to +\infty} |\partial \Lambda_n|/ |\Lambda_n| =0$.
An example of such a sequence can be the sequence of cubes $\Lambda_{L_n}$ for any strictly increasing sequence of positive integers $L_n$.
For a van Hove sequence $\{\Lambda_n\}_{n\in \mathbb{N}}$, we denote
\begin{equation}
  \label{q8}
F(h) = \lim_{n\to +\infty }F_{\Lambda_n} (h), \quad F^\mu(h) = \lim_{n\to +\infty }F^\mu_{\Lambda_n} (h),
\end{equation}
where $F_{\Lambda_n} (h)$ and $F^\mu_{\Lambda_n} (h)$ are as in (\ref{q5}) and (\ref{q7}), respectively. Furthermore, for a strictly increasing sequence $\{L_n\}$ and $F_{L_n}(h)$ as in (\ref{q6}), we set
\begin{equation}
  \label{q9}
F^{\rm per} (h) = \lim_{n\to +\infty}F_{L_n}(h).
\end{equation}
It is known, see Theorem 5.1.3 on page 268 in \cite{[A3]},  that, for any $\mu\in \mathcal{G}$ and for any van Hove sequence
$\{\Lambda_n\}_{n\in \mathbb{N}}$, both limits in (\ref{q8}) exist, do not depend on the choice of $\{\Lambda_n\}_{n\in \mathbb{N}}$, and are equal to each other. Furthermore, for any sequence
$\{L_n\}_{n\in \mathbb{N}}$ such that $L_n \to +\infty$, the limit in (\ref{q9}) exists, is independent of the choice of
$\{L_n\}_{n\in \mathbb{N}}$, and is equal to the limits in (\ref{q8}). That is,
\begin{equation}
\label{q9a}
F^{\rm per} (h) = F(h) = F^\mu (h), \qquad {\rm for} \ \ {\rm all} \ \ \mu\in \mathcal{G}.
\end{equation}
Therefore, $F(h)$ is a universal thermodynamic function characterizing the model, that along with $h$ depends also on $\beta$. Note
that in \cite{[A3]}, instead of $F(h)$  we considered the {\it pressure} $p(h) = - \beta F(h)$.

We recall that the model (\ref{1}) with $\nu>1$ is considered only with $V$ as in (i). In this case the free energy density depends only on
the norm of the external field, which we can choose therefore in the form $(h, 0, \dots , 0)$ with $h\in \mathbb{R}$. In the remaining (ii) and (iii) cases, we have $\nu=1$ and hence $h\in \mathbb{R}$.
Thus, from now on we assume that $h\in \mathbb{R}$ in all cases.
By (\ref{q5}), (\ref{11z}), and (\ref{q2}), we obtain
\begin{eqnarray}
  \label{q10}
- \frac{\partial}{\partial h} F_\Lambda (h) & = &  - \frac{1}{\beta |\Lambda|}\frac{\partial}{\partial h} Z_\Lambda\\[.2cm] & = & \frac{1}{ |\Lambda|} \sum_{x\in \Lambda} \varrho_\Lambda [q^{(1)}_x ]=: M_\Lambda (h). \nonumber
\end{eqnarray}
Here $M_\Lambda (h)$ is  the {\it local  polarization} averaged over $\Lambda$. Likewise,
\begin{eqnarray}
  \label{q11}
-\frac{\partial}{\partial h} F_L (h) & = & - \frac{1}{\beta |\Lambda|}\frac{\partial}{\partial h} Z_L\\[.2cm] & = & \frac{1}{ |\Lambda|} \sum_{x\in \Lambda_L} \varrho_L [q^{(1)}_x ]= \varrho_L [q^{(1)}_x ]=:  M_{L} (h). \nonumber
\end{eqnarray}
Here we have taken into account that the periodic state $\varrho_L$ is invariant with respect to translations of the torus $\Lambda_L$.  The formulas just obtained relate the local free energy densities with the
local polarizations. However, the convergence in (\ref{q8}) and (\ref{q9}) need not yield the convergence of the derivatives, i.e., of the local polarizations.
Computing  one more derivative in (\ref{q10}) we obtain, see page 220 in \cite{[A3]},
\begin{widetext}
\begin{equation}
  \label{q12}
-\frac{\partial^2}{\partial h^2} F_\Lambda (h)  =  \frac{1}{2\beta |\Lambda|}\int_{\Omega_\Lambda}\int_{\Omega_\Lambda}
\left[\sum_{x\in \Lambda}\int_0^\beta\left(\omega_x(\tau) - \tilde{\omega}_x(\tau) \right)d \tau \right]^2 \nu_\Lambda (d \omega_\Lambda) \nu_\Lambda (d \tilde{\omega}_\Lambda) \geq 0 .
\end{equation}
\end{widetext}
The latter inequality follows from the fact that we integrate a nonnegative expression, i.e., $\left[ \cdots \right]^2$. From this inequality we see that
 $F_\Lambda$ is a concave function of $h$. The same is true also for $F_L$. Thus, the limiting
free energy density given in (\ref{q8}) is also a concave function of $h$. Concave functions always have one-sided
derivatives
\begin{equation}
  \label{q13}
  M_{\pm } (h) = - \lim_{\varepsilon \to 0+} \frac{F(h \pm \varepsilon) - F(h)}{\pm \varepsilon}.
\end{equation}
If $M_{+} (h) = M_{-}(h)$, then $F(h)$ is differentiable at this $h$, and one can speak of the {\it global
polarization}
\begin{equation}
  \label{q14}
M(h) := M_{+} (h) = M_{-} (h) = - F'(h).
\end{equation}
Given $h_*\in \mathbb{R}$, we say that $M(h_*)$ does not exist if $M_{+} (h_*) \neq M_{-}(h_*)$.
By the concavity of $F$, if the global polarization $M(h)$ exists for a given $h$, it can be obtained
as a limit of the sequences of local polarizations
 (\ref{q10}), (\ref{q11}), which is independent of the choice of
the corresponding van Hove sequences $\{\Lambda_n\}$ or $\{L_n\}$. That is,
\begin{equation}
  \label{q15}
 M(h) = \lim_{n\to +\infty} M_{\Lambda_n} (h) = \lim_{n\to +\infty} M_{L_n} (h).
\end{equation}
Suppose now that,
for a given $h_*\in \mathbb{R}$, $F(h)$ is not differentiable at $h_*$, but is differentiable on the intervals
$(h_* - \varepsilon, h_*)$ and $(h_* , h_*+ \varepsilon)$ for some $\varepsilon >0$. Then $M_{+} (h_*) \neq M_{-} (h_*)$, i.e., the polarization $M(h)$
is discontinuous (makes a jump) at $h_*$, which physicists interpret as a phase transition. However, we failed so far to prove that
the discontinuity of $M(h)$ implies the multiplicity of phases, i.e., $|{\rm ex}(\mathcal{G})| >1$, even for the simplest versions of
the model\footnote{Such a statement holds
true for the Ising model, see Theorem III.3.11, page 260 in \cite{[Simon]}, as well as for the model of a nonideal gas in $\mathbb{R}^d$, see \cite{[Klein]}.} (\ref{1}).
The converse statement holds true in the following form, see Theorem 5.3.3, page 276 in \cite{[A3]}.
\begin{theorem}
  \label{Ft}
  For $\nu=1$, if the free energy density is differentiable in $h$ at a given $h_*$ and $\beta$, then the set of Gibbs measures $\mathcal{G}$ is a singleton,
and hence there exists only one thermodynamic phase at these $\beta$ and $h_*$.
\end{theorem}
In the next subsection, we return to the connection of the properties of $F(h)$
with phase transitions.

We conclude this subsection  with the following remark.
In some cases, it is possible to find
values of $h$
where $F(h)$ is differentiable for all $\beta$. In particular, see  Theorem 5.2.3 on page 273 in \cite{[A3]},
if $\nu = 1,2$ and $V$ is as in (\ref{3}), then $F(h)$ is infinitely differentiable at all $h\neq 0$, and hence $M(h)$ exists at all such $h$. This result follows
from a generalization of the celebrated Lee-Yang theorem, see also Theorem 5.2.4 on page 275 in \cite{[A3]}.

\subsection{Phase transitions and order parameters}

\label{SS2.5}

We recall that according to our definition a phase transition in the model (\ref{1}) occurs  if it
has multiple thermodynamic phases existing at the same values of $\beta$ and $h$. Mathematically,
this means that the set of Gibbs measures $\mathcal{G}$, and hence its extreme boundary ${\rm ex}(\mathcal{G})$,
contain more than one element. So far, for models like (\ref{1})
there exists only one way of establishing the latter fact. Namely, one shows the existence of an element of
$\mathcal{G}$, which is not extreme (not a pure state). Recall that ${\rm ex}(\mathcal{G})$ is a subset of  $\mathcal{G}$.
A candidate to be such an element is the so called `periodic state' which one obtains as the limit of the measures
(\ref{10a}) as $L\to +\infty$. For $h=0$ and big enough $\beta$, this state can be {\it nonergodic} with respect to the translations of $\mathbb{Z}^d$, whereas extreme states are always ergodic.
We use this argument in the next section. However, this way essentially employs the $O(\nu)$-symmetry,  and hence
does not cover case (iii). Therefore, to mathematically describe phase transitions in a wider class of models one needs other definitions,
 consistent with the physical point of view on this subject. In this subsection, we present and analyze such alternative definitions.

The first definition is based on the celebrated L. D. Landau classification, which employs the differentiability of the free energy density in $h$,
see, e.g., Chapter I in \cite{[BC]}. Of course, this property depends also on the value of $\beta$.
Namely, we say that the model (\ref{1}) has a {\it first-order} phase transition at certain values of $\beta$ and $h_*$ if at this $\beta$,  the polarization $M(h)$ is discontinuous
at $h_*$. The model has a {\it second-order} phase transition at $\beta$ and $h_*$ if the first derivative $F'(h)$ is continuous, but the second derivative $F''(h)$
is discontinuous at $h_*$. Note that these definitions can be applied to version (iii) of our model.

The second definition employs an {\it order parameter}. It can be applied to cases (i) and (ii) of the model (\ref{1}) with $h=0$,
which we suppose to be $O(\nu)$-symmetric. One can  show that the local states (\ref{5}) and (\ref{5a}) are such that
$\varrho_\Lambda (|q_x|^2)$ and  $\varrho_L (|q_x|^2)$ exist and are finite, in spite of the fact that $|q_x|^2$ is an unbounded operator. Then we can consider the Matsubara function, cf. (\ref{8z}) and (\ref{204}),
\begin{eqnarray}
\label{q16}
& & \Gamma^L_{xy} (\tau, \tau')\\[.2cm] & & \quad = \frac{1}{Z_L}\sum_{j=1}^\nu {\rm trace}\left[q^{(j)}_x e^{- (\tau'-\tau)H_L} q^{(j)}_y e^{- (\beta +\tau-\tau')H_L} \right]. \nonumber
\end{eqnarray}
Here $0\leq \tau \leq \tau' \leq \beta$, $x$ and $y$ are points in the cube $\Lambda_L$, $L\in \mathbb{N}$, and $H_L$ and $Z_L$ are given in (\ref{4b}) and (\ref{4c}), respectively.
According to (\ref{9}) we have
\begin{equation}
  \label{q17}
\Gamma^L_{xy} (\tau, \tau') = \sum_{j=1}^\nu \int_{\Omega_{\Lambda_L}} \omega_x^{(j)} (\tau)  \omega_y^{(j)} (\tau') \nu_L (d
\omega_{\Lambda_L}),
\end{equation}
where $\nu_L$ is given in  (\ref{10a}). By the Cauchy-Schwarz inequality, and then by a property of
the integrals with respect to $\nu_L$, it follows that
\begin{equation}
  \label{q17a}
  \left\vert \Gamma^L_{xy} (\tau, \tau') \right\vert \leq \varrho_L (|q_x|^2) \leq C.
\end{equation}
Note that $\varrho_L (|q_x|^2)$ is independent of $x$. The meaning of the second inequality in (\ref{q17a})
is that the bound $C$ is independent of $\Lambda_L$.

As a function of $\tau$ and $\tau'$, $\Gamma^L_{xy}$ in (\ref{q17}) can clearly be extended to all values of $(\tau, \tau') \in [0,\beta]^2$.
This extension is a continuous function of
\begin{equation*}
|\tau - \tau'|_\beta := \min\{ |\tau-\tau'|; \beta -  |\tau-\tau'|\}.
\end{equation*}
Then the following integral makes sense
\begin{equation}
  \label{q19}
 D^L_{xy} = \beta^{-1} \int_0^\beta \int_0^\beta \Gamma^L_{xy} (\tau, \tau') d\tau d \tau'.
\end{equation}
If the model (\ref{1}) is $O(\nu)$-symmetric, as it is in cases (i) and (ii) with $h=0$,
$D^L_{xy}$ describes static correlations between the oscillators located at $x$ and $y$. In view of the
translation symmetry of $\nu_L$, $D^L_{xy}$ is a function of the periodic distance $|x-y|_L$, see (\ref{4z}). Now we introduce, cf. (\ref{q17}),
\begin{eqnarray}
  \label{q20}
P_L & = & \frac{1}{\beta |\Lambda_L|^2} \sum_{x, y \in \Lambda_L} D^L_{xy} \\[.2cm]
& = & \int_{\Omega_{\Lambda_L}}\left\vert\frac{1}{\beta|\Lambda_L|}\sum_{j=1}^\nu \sum_{x\in \Lambda_L} \int_0^\beta \omega^{(j)}_x (\tau) d \tau\right\vert^2 \nu_L (d \omega_{\Lambda_L}). \nonumber
\end{eqnarray}
If $D^L_{xy}$ decays sufficiently fast as $|x-y|_L \to +\infty$, then the double sum in the first line of (\ref{q20}) is of order strictly less than $|\Lambda_L|^2$,
and hence $P_L \to 0$ as $L\to +\infty$. On the other hand, by means of (\ref{q17a}) we readily obtain that $P_L \leq C$, which means that the sequence $\{P_L\}_{L\in \mathbb{N}}$ is bounded and hence contains convergent
subsequences. The largest limit of such subsequences, i.e.,
\begin{equation}
  \label{q21}
 P = \limsup_{L\rightarrow +\infty} P_L,
\end{equation}
is called {\it order parameter}.  We say that, at a given value of $\beta$, there exists a {\it long range order,}
if $P>0$ for this $\beta$. In this case, we also say that the model is in an {\it ordered state}. Thus, we have the following three definitions
of a phase transition:
\vskip.2cm
\begin{tabular}{ll}
(a) \ &employing thermodynamic phases: $|{\rm ex}(\mathcal{G})|>1$;\\[.2cm]
(b) \ &employing the polarization jump:\\[.1cm] &\qquad \qquad $M_{+} (h) \neq M_{-}(h)$;\\[.2cm]
(c) \ &employing  the order parameter: $P>0$.
\end{tabular}
\vskip.2cm \noindent
The connections between these notions will be analyzed in the next section.

\subsection{Infrared bounds}
\label{SS2.6}

The fact that $D^L_{xy}$ is a function of the periodic distance  (\ref{4z}) allows us to use the Fourier transformation
in the torus $\Lambda_L$. Let $\Lambda^*_L$ denote the Brillouin zone for $\Lambda_L$, that is, the set consisting of $p= (p_1 , \dots p_d)$ such that
\[
p_i =-\pi+ \pi s_i/L, \quad s_i = 1,2, \dots , 2L, \quad i=1, \dots , d.
\]
By $(-\pi, \pi]^d$ we denote the set $$\{p= (p_1 , \dots, p_d): p_i \in (-\pi, \pi]\},$$
which is the Brillouin zone for the whole crystal.
For $p\in (-\pi, \pi]^d$, we define
\begin{equation}
  \label{q22}
\widehat{D}^L_p = \sum_{y\in \Lambda_L} D^L_{xy} \exp({\rm i} (p, x-y)).
\end{equation}
The latter sum is finite for finite $L$, but can be divergent in the limit $L\to +\infty$.
Traditionally,  this is called the {\it infrared divergence}. The restriction of $\widehat{D}^L_p$ to
$\Lambda^*_L$ can be used to define the inverse transform
\begin{eqnarray*}
  D^L_{xy} & = & \frac{1}{|\Lambda_L|} \sum_{p\in \Lambda^*_L} \widehat{D}^L_p \exp(-{\rm i} (p, x-y))\\[.2cm] & = & \frac{1}{|\Lambda_L|} \sum_{p\in \Lambda^*_L} \widehat{D}^L_p \cos(p, x-y). \nonumber
\end{eqnarray*}
Suppose now that we are given a continuous function $\widehat{B}: (-\pi, \pi]^d \rightarrow (0, +\infty]$ with the following properties:
\begin{eqnarray}
  \label{q24}
 & & {\rm (a)} \quad \int_{ (-\pi, \pi]^d} \widehat{B}(p) d p < +\infty; \\[.4cm]
 & &  {\rm (b)} \quad  \widehat{D}^L_p \leq \widehat{B}(p), \qquad {\rm for} \ \ {\rm all} \ \ p \neq 0. \nonumber
\end{eqnarray}
Since $\widehat{B}$ is independent of $L$, by (b) one can control the infrared divergence mentioned above. That is why (b) is called
an {\it infrared bound}, see \cite{[KK1]} for more detail on this item.
In view of (a), for $x,y \in \mathbb{Z}^d$ the following quantity
\begin{equation}
  \label{q25}
B_{xy} = \frac{1}{(2\pi)^d} \int_{(-\pi, \pi]^d} \widehat{B}(p)\cos(p,x-y) d p
\end{equation}
is well-defined. For $x,y \in \Lambda_L,$ we also set
\begin{equation*}
 B^L_{xy} = \frac{1}{ |\Lambda_L|} \sum_{p\in \Lambda^*_L, \ p\neq 0}\widehat{B}(p)\cos(p,x-y) ,
\end{equation*}
and $B^L_{xy}=0$ if either of $x,y$ belongs to $\Lambda_L^c$. One can prove, see Proposition 6.1.2 on page 284 in \cite{[A3]}, that, for every $x$ and $y$,
$B^L_{xy} \rightarrow B_{xy}$ as $L \to +\infty$. Furthermore, see Lemma 6.1.3 {\it ibid}, for every  $\Lambda_L$ and any  $x,y \in \Lambda_L$, under the conditions in (\ref{q24}) the following holds
\begin{equation}
  \label{q27}
  D^L_{xy} \geq \left( D^L_{xx} - B^L_{xx}\right) + B^L_{xy}.
\end{equation}
Note that both $D^L_{xx}$ and $B^L_{xx}$ are independent of $x$.
The meaning of (\ref{q27}) is that the infrared bound (b) in (\ref{q24}) allows one to control from below the decay of the static correlations.
Suppose now that there exists a positive $\vartheta$ such that,
for any cube $\Lambda_L$,
\begin{equation}
  \label{q28}
D^L_{xx} \geq \vartheta.
\end{equation}
Suppose also that (\ref{q24}) holds and that $B_{xy}$ obeys the following conditions:
\begin{eqnarray}
  \label{q29}
 & & {\rm (a)} \quad \vartheta > B_{xx}; \\[.3cm]
 & & {\rm (b)} \quad \lim_{|x-y|\to +\infty} B_{xy} = 0. \nonumber
\end{eqnarray}
If this is the case, for some $\varepsilon >0$, all $\Lambda_L$ and all  $x,y\in \Lambda_L$ such that $|x-y|\geq \delta(\varepsilon)$,
by (\ref{q27}) we get that $D^L_{xy} \geq \varepsilon$, which yields in (\ref{q20}) that $P_L \geq \varepsilon/\beta$ and hence $P>0$, see (\ref{q21}). Thus, (\ref{q28}) and (\ref{q29}) imply the existence of the long range order mentioned above.

It turns out that, for our model (\ref{1}) with any value of the dimension $\nu \in \mathbb{N}$, and with any $V$,
 the function $\widehat{B}(p)$ can be found explicitly.
The only conditions are that the lattice dimension $d$ should be at least $3$ and the interaction in (\ref{1}) should be of nearest neighbor type, i.e., $J_{xy}=J>0$ whenever $|x-y|=1$, and
$J_{xy}=0$ otherwise. Then this function has the following form, see Corollary 6.2.9, page 303 in \cite{[A3]},
\begin{eqnarray}
\label{q30}
 & & \quad  \widehat{B}(p)=  \frac{\nu}{2 J \varepsilon (p)}, \\[.2cm] & & \varepsilon (p)   :=  \sum_{i=1}^d (1 - \cos p_i). \nonumber
\end{eqnarray}
Note  that  $\widehat{B}(p)\sim C |p|^{-2}$, as $|p| \to 0$. Hence, it satisfies condition (a) in (\ref{q24})
only for $d\geq3$. By (\ref{q25}), we get
\begin{eqnarray}
  \label{q31}
& & \qquad \ B_{xx} =  \nu \theta (d) / 2 J,  \\[.2cm] & &  \theta (d) := \frac{1}{(2 \pi)^d}\int_{(-\pi, \pi]^d} \frac{dp}{\varepsilon (p)} . \nonumber
\end{eqnarray}
Then, for any $\vartheta>0$, condition (a) in (\ref{q29}) can be satisfied by taking big enough $J$.
The validity of (b) in (\ref{q29}) follows by the Riemann-Lebesgue lemma, see Proposition 6.2.10, page 303 in \cite{[A3]}.

\section{The Results}
\label{SS3}

\subsection{Phase transitions}
\label{SS3.1}

In the first subsection below, we analyze the relationships between the three definitions of a phase transition given above,
and then present the corresponding  sufficient conditions for the particular versions of our model.
\subsubsection{The order parameter and the first-order phase transition}
First of all we note that the order parameter (\ref{q21}) can be defined without using any path integrals, see
(\ref{q16}), (\ref{q19}), and (\ref{q20}). The same is true for the free energy density (\ref{q9}) since
\[ F_{L} (h) = \frac{1}{\beta |\Lambda_L|} \left(\ln Z^{\rm har}_{L} - \ln Z_L \right),
\]
see (\ref{q6}), (\ref{q4}), and (\ref{4c}). However, the path integral representation of these quantities allows us to
apply here measure-theoretic tools, which proved to be useful in classical statistical physics. Namely, by means of
Griffiths' theorem,  see Theorem 6.1.7, page 286 in \cite{[A3]}, we obtain the following result.
\begin{theorem}
  \label{1tm}
 Let the dimension $\nu$ be arbitrary and the potential $V$ be $O(\nu)$-symmetric and such that the estimate
(\ref{3ba}) holds. Then the one-sided derivative $M_{+}(0)$, see (\ref{q13}), and the order parameter $P$, see (\ref{q21}), obey the estimate
\begin{equation}
  \label{R1}
  M_{+} (0) \geq \sqrt{P}.
\end{equation}
Hence, if the model with such $V$ is in an ordered state, then it undergoes a first-order
phase transition.
\end{theorem}
Indeed, by the assumed symmetry we have $M_{+} (0) = - M_{-} (0)$. Hence, $P>0$  by (\ref{R1}) yields  $M_{+} (0) >0$, and then $M_{-} (0)<0$.
Note, however, that the positivity of the order parameter need not yet mean that the thermodynamic phases are multiple. To
prove that this is the case we have to impose further restrictions on $\nu$ and $V$.

\subsubsection{Phase transition in the $|q|^4$ case}

Here we suppose that $\nu$ is arbitrary and $V$ is as in (\ref{3}). Recall that $a>0$ is the rigidity parameter, see (\ref{2}).
Then, for $\alpha$ and $b$ as in (\ref{3}), we set
\begin{equation}
  \label{R2}
\vartheta_* = \frac{\nu\alpha}{2b (\nu+2)}.
\end{equation}
Note that $\vartheta_*>0$ whenever $\alpha>0$ and thereby the potential energy in (\ref{2}) has multiple minima.
For $\nu=1$, $V(q)$ has two minima at $\pm q_*$, with $q_* = \sqrt{3\vartheta_*}$, cf. (\ref{00}).

Let $f:[0,+\infty) \to (0, 1]$  be defined  as
\[
 f(0) = 1, \ \ \ {\rm and} \ \ \ f ( u \tanh u) = u^{-1} \tanh u, \ \ {\rm for} \ \ u>0.
\]
Then, for a fixed $\gamma>0$,  the function
\begin{equation}
  \label{15}
 \phi(t) := t \gamma  f(t/\gamma), \quad t>0
\end{equation}
is differentiable and monotone increasing to $\gamma^2$ as $t \to +\infty$, that readily follows from the definition of $f$.
Thus, we set $\gamma = 4 m \vartheta_*$ and obtain that the function
\[
\frac{1}{2m} J \phi(\beta) = 2 \beta J \vartheta_* f(\beta/ 4 m \vartheta)
\]
increases as  $\beta \to + \infty$ and tends to $8 m \vartheta_*^2 J$. Therefore, if the following condition holds
\begin{equation}
 \label{17}
8 m \vartheta_*^2 J > \theta (d),
\end{equation}
then the equation
\begin{equation}
  \label{18}
   2 \beta J \vartheta_* f(\beta/ 4 m \vartheta_*) = \theta(d).
\end{equation}
has a unique solution, which we denote by $\beta_*$.
The following statement,  see Theorem 6.3.6, page 308 in \cite{[A3]}, gives a sufficient condition
for a phase transition to occur.
\begin{theorem}
  \label{2tm}
Let $d\geq 3$, the interaction intensity be $J_{xy} = J>0$ whenever $|x-y|=1$, and $J_{xy}=0$ otherwise, and let
$V$ be as in (\ref{3}). Suppose also that the condition (\ref{17}) is satisfied, and hence (\ref{18}) has the solution $\beta_*$.
Then, for $\beta > \beta_*$,
the following holds: (a) $|{\rm ex}( \mathcal{G})|>1$; (b) $P>0$; (c) $M_{+}(0) > M_{-}(0)$.
\end{theorem}
The implication $(b) \Rightarrow (c)$ follows by Theorem \ref{1tm}.
The proof of (a) and (b) relies on showing that the estimates (\ref{q28}) and (\ref{q29}) hold with
$B_{xy}$ as in (\ref{q30}) and (\ref{q31}). First, for the state (\ref{5a}) and the Hamiltonian (\ref{4b}), and for an appropriate
operator $A$, one readily gets that
\[
\varrho_L\left(\left[A ,\left[H_L,A \right] \right] \right) \geq 0,
\]
where $[H_L , A] = H_L A - A H_L$. Applying this formula with $A= p^{(j)}_x$, $x\in \Lambda_L$ and $j= 1 , \dots , \nu$, and taking
into account the form of $V$ and the commutation relations (\ref{3dd}), we obtain
\begin{equation}
  \label{R3}
  \varrho_L \left( |q_x|^2\right) \geq \vartheta_*.
\end{equation}
Next, we employ the Bruch-Falk inequality, see also \cite{[DLS]} and page 392 in \cite{[Simon]}, and obtain from the latter that
\begin{equation}
  \label{R4}
  D^L_{xx} \geq \beta \nu \vartheta_* f\left(\frac{\beta}{4 m \vartheta_*}\right),
\end{equation}
where $f$ is as in (\ref{15}). This estimate and (\ref{q31}) yield
\begin{equation}
  \label{R5}
D^L_{xx} - B_{xx} \geq   \beta \nu \vartheta_* f\left(\frac{\beta}{4 m \vartheta_*}\right) - \frac{\nu \theta (d)}{ 2J}.
\end{equation}
For $\beta > \beta_*$, the right-hand side of (\ref{R5}) is strictly positive. Then $D^L_{xy}$ does not decay to zero as
$L\to +\infty$ and $|x-y|_{L} \to +\infty$. This yields $P>0$, see (\ref{q20}). This also yields that the periodic state
is not ergodic with respect to the group of translations of $\mathbb{Z}^d$, implying that $\mathcal{G}$ contains elements which do not belong to ${\rm ex}(\mathcal{G})$,  see Definition 3.1.26 and Corollary 3.1.29, page 207 in \cite{[A3]}. By this we get that
$|\mathcal{G}|>1$, which proves also {\it (a)}.

\subsubsection{Phase transition in the $\nu =1$ symmetric case}

Here we assume that $V(q) = V(-q)$ and (\ref{3a}) is satisfied. In addition, we assume that there exists $J>0$ such that, for all $x,y \in \mathbb{Z}^d$ obeying $|x-y|=1$, the following holds
\begin{equation}
  \label{R6}
 J_{xy} \geq J .
\end{equation}
Let $\Phi$ be as in (\ref{3c}). Then the equation
\begin{equation}
  \label{R7}
 - 2 b_1  + \Phi(\vartheta) = 0,
\end{equation}
has a unique solution $\vartheta_* >0$, since $b_1 >0$, see (\ref{3b}). In this case, we have the following result, see
Theorem 6.3.8, page 310 in \cite{[A3]}.
\begin{theorem}
  \label{Rtm}
Let the model be as just described and let the condition in (\ref{17}) be satisfied with $\vartheta_*$ defined by (\ref{R7}) and $J$ obeying (\ref{R6}). Then, for every $\beta>\beta_*$ defined by (\ref{18}), the following holds:  (a) $|{\rm ex}( \mathcal{G})|>1$; (b) $P>0$; (c) $M_{+}(0) > M_{-}(0)$.
\end{theorem}
Again, as above $(b)$ implies $(c)$ by Theorem \ref{1tm}. Claims $(a)$ and $(b)$ are proven by comparing the considered model
with the model described by the Hamiltonian
\begin{equation}
  \label{R8}
H^{\rm low} = - \frac{J}{2}\sum_{x,y: \ |x-y|=1} q_x q_y + \sum_{x} \left[\frac{1}{2m}p_x^2 + W(q_x) \right],
\end{equation}
where $W$ is as in (\ref{3b}). Then we use the  GKS inequalities, which hold in the considered case, see Theorem 2.2.2, page 163 in \cite{[A3]},
and a comparison method developed in \cite{[KP]}. By means of these tools we prove that if $(a)$ and $(b)$ hold for the model as in (\ref{R8}), they hold also for that of (\ref{1}) with such $V$ and $\nu=1$. On the other hand, the properties of $W$ allow us to prove that (\ref{R4}) holds for
the model (\ref{R8}) with $\vartheta_*$ defined by (\ref{R7}). The infrared bound (\ref{q22}) with $\widehat{B}_p$ as in (\ref{q30}) also holds
since the interaction is of nearest-neighbor type. Then the proof of $(a)$ and $(b)$ for the model (\ref{R8}) follows as for Theorem \ref{2tm}.

\subsubsection{Phase transition in the $\nu =1$ asymmetric case}

In case (iii) of the model (\ref{1}), the only result concerning phase transitions which we managed to get so far is
a statement that the model parameters and the inverse temperature $\beta$ can be chosen in such a way that the polarization $M(h)$ becomes discontinuous
at certain $h_*$, i.e., the model has a first-order phase transition, see Theorem 3.4 of \cite{[KKK]} and  Section 6.3.3 of \cite{[A3]}.
\begin{theorem}
  \label{R2tm}
Suppose that $\nu=1$, $d\geq 3$, the interaction is of nearest-neighbor type, and $V$ is continuous and such that (\ref{3ba}) holds, which includes the case of $V(q) \neq V(-q)$. Then, for every $\beta >0$, there
 exist $m_*>0$ and $J_*>0$ such that, for all $m>m_*$ and $J>J_*$, there exists $h_*\in \mathbb{R}$, possibly dependent on $m$, $\beta$, and $J$,
such that the polarization $M(h)$ becomes discontinuous at $h=h_*$, i.e., the model has a first-order phase transition.
\end{theorem}
 Here we again exploit the fact, used in the proof of Theorem \ref{2tm}, that the existence of a nonergodic state implies $|\mathcal{G}|=+\infty$, and hence a first-order phase transition. Recall that the existence of a nonergodic
 periodic  state would follow from the fact that the static
correlation function $D^L_{xy} - [M_L(h)]^2$ does not decay to zero as $L\to +\infty$ and $|x-y|_L \to +\infty$. The latter means that there exists
 a sequence of integer numbers $\{L_n\}_{n\in \mathbb{N}}$ such that $L_n \to +\infty$ and the following holds
\begin{equation}
  \label{D}
 D^{L_n}_{xy} - [M_{L_n}(h)]^2 \geq \delta,
\end{equation}
for some $\delta >0$, all $x,y$, and all $n$ such that $x,y \in \Lambda_{L_n}$.
Note that in the symmetric case, we have $M_L(0)=0$,  and hence (\ref{q27}) can be used.
Suppose now that there exists $h_0$ such that $M(h_0)=0$. Since $M(h_0) = \lim M_L (h_0)$, see (\ref{q15}), for such $h_0$ (\ref{D}) would follow from the fact that
\begin{equation}
  \label{D1}
  D^{L_n}_{xy} \geq \delta.
\end{equation}
 Thus, if (\ref{D1}) holds and $M(h)$ takes both negative and positive values, then
either $M(h_0)=0$, and thus (\ref{D}) yields a first-order phase transition, or $M(h)$ is discontinuous, which yields the same.

To realize this scheme we crucially use the properties of our path integrals.
As was mentioned above, the free energy density $F(h)$ is a concave function of $h\in \mathbb{R}$, see (\ref{q12}). Then the set of such $h$ where $M_{+}(h) \neq M_{-}(h)$, and hence $M(h)$ does not exist, is at most countable.
Let $\mathcal{R}$  denote the set of all $h\in \mathbb{R}$ for which $M(h)$ exists.
Then any interval $(a,b) \subset \mathbb{R}$ contains points of $\mathcal{R}$.
A tedious analysis of the properties of the path integrals used in our constructions
yields the following estimate of the free energy density (\ref{q6})
\begin{equation}
  \label{R9}
 - F_L (h) \geq h\varepsilon + \Psi(\beta ) + \Xi (m),
\end{equation}
see Theorem 5.2.2, page 271 in \cite{[A3]}. This estimate
holds for all $h\geq 0$,  $J\geq 0$, and $m>m_0$, where $m_0>0$ is a certain quantity which may depend on $\beta$. Here $\varepsilon>0$ is a constant, and $\Psi$ and $\Xi$ are certain real-valued continuous functions of $\beta$ and $m$, which can be calculated explicitly. As the right-hand side of (\ref{R9}) is independent of $L$, this estimate holds also in the limit $L\to +\infty$,
for all $h$ such that $M(h)$ exists. By the mentioned concavity of $F(h)$, for any such $h\geq 0$, we have
\begin{eqnarray*}
  M(h) & \geq & \left[F(0) - F(h) \right]/h\\[.2cm] & \geq & \varepsilon + \left[F(0) +  \Psi(\beta ) + \Xi (m) \right]/h.
\end{eqnarray*}
Since the right-hand side of the latter inequality tends to $\varepsilon >0$ as $h\to + \infty$, $M(h)$ becomes positive for big enough $h$. Thus, there exists $h_{+}>0$ such that $M(h) >0$ for all $h\in \mathcal{R}\cap (h_{+}, +\infty)$. Likewise, one can show that
 there exists $h_{-} < 0$ such that $M(h) <0$ for all $h\in \mathcal{R}\cap(-\infty, h_{-})$.
 Next, some additional analysis
of the path integral in the representation
\[
\varrho_L \left(q_x^2 \right) = \int_{\Omega_{\Lambda_L}} \left[ \omega_x (0)\right]^2 \nu_L (d \omega_{\Lambda_L}),
\]
cf. (\ref{10a}),
yields, see Lemma 6.3.9, page 312 in \cite{[A3]}, that, for every positive $\beta$ and $\vartheta_*$, there exist
positive $m_1$ and $J_*$, which may depend on $\beta$ and $\vartheta_*$ but are independent of $h$, such that,
for any cube $\Lambda_L$ and any $h\in \mathbb{R}$, and for all $J>J_*$ and $m> m_1$, the following holds
\begin{equation}
  \label{R10}
  \varrho_L \left(q_x^2 \right) \geq \vartheta_*.
\end{equation}
Then the proof of Theorem \ref{R2tm} follows along the next line of arguments. Fix any $\vartheta_*$ and take $m> m_*:=\max\{m_0; m_1\}$. Then both (\ref{R9}) and (\ref{R10}) hold. Hence, in view of (\ref{R10}), we have that (\ref{R4}) and (\ref{R5}) hold for such $\beta$ and $m$, and for all $L$ and $h$. Then we increase $J$, if necessary, up to the value at which the right-hand side of (\ref{R5}) gets strictly positive. Afterwards, all the parameters, except for $h$, are kept fixed. By (\ref{q27}), the positivity of the RHS of (\ref{R5})  yields (\ref{D1}). If $M(h)$ were everywhere continuous, the fact that it takes both positive and negative values would imply that
there exists $h_0$ such that $M(h_0)=0$, which together with (\ref{D1}) would yield (\ref{D}). The latter, however, contradicts the everywhere continuity of $M$, see Theorem \ref{Ft}.

\subsection{Quantum effects}
\label{SS3.2}
\subsubsection{The stability of quantum crystals}
To understand what makes a crystal stable or unstable let us first consider the scalar ($\nu=1$) harmonic version of the model
(\ref{1}), in which $H_x=H_x^{\rm har}$, see (\ref{2}). For this model, the global Gibbs states, and hence the thermodynamic phases as mathematical objects,  can be constructed only if the stability condition
\begin{equation*}
  \hat{J}_0 < a = m [\Delta^{\rm har}]^2
\end{equation*}
is satisfied, see (\ref{J}) and (\ref{V3}). In this case, $|\mathcal{G}|=1$ for all $\beta$, see also Theorem \ref{3tm} below . If $\hat{J}_0 =a$, the harmonic crystal becomes unstable
with respect to spatial translations; for $d\geq 3$, the set $\mathcal{G}$ is still a singleton, and $\mathcal{G}= \emptyset$ for $d\leq 2$, because
of the divergence of integrals as in (\ref{q31}) for such values of $d$.
 In the anharmonic case, due to the assumption (\ref{3ba}) global Gibbs states exist for all $\hat{J}_0 >0$, and the
instability of the crystal can be caused by the `effective' change of the equilibrium positions of oscillators, i.e., by a structural phase transition.
Thus, according to Theorem \ref{1}, a sufficient condition for a $d$-dimensional quantum crystal, $d\geq 3$, to have multiple thermodynamic phases at a given $\beta$ is
\begin{equation}
  \label{S2}
 \beta \hat{J}_0 \vartheta_* f \left( \frac{\beta}{4 m \vartheta_*}\right) > d \theta (d),
\end{equation}
which in the classical limit $m \to +\infty$, cf. (\ref{2a}), takes the form
\begin{equation*}
  \beta \hat{J}_0 \vartheta_* >  d \theta (d).
\end{equation*}
The latter condition can be satisfied by picking big enough $\beta$. Therefore, the classical anharmonic crystal with $d\geq 3$ always has a
phase transition - no matter how small $\hat{J}_0$ is. For finite $m$, the left-hand side of (\ref{S2}) is bounded by $4m \vartheta_*^2 \hat{J}_0$,
and the bound is achieved in the limit $\beta \to + \infty$. Thus, if
\begin{equation}
  \label{S4}
Q:= 4m \vartheta_*^2 \hat{J}_0 \leq  d \theta (d),
\end{equation}
the condition (\ref{S2}) will not be satisfied for any $\beta>0$. Although (\ref{S2}) is only a sufficient condition, one might expect that
the phase transition cannot occur at any $\beta$ if the parameter $Q$ is small. This effect could be called {\it quantum stabilization}. The following statement shows that this  really occurs.
\begin{theorem}
 \label{3tm}
Let $\nu=1$ and the potential $V$ be as in (\ref{V1}). Then the set $\mathcal{G}$ of the Gibbs states of the corresponding model is a singleton at all $\beta>0$ under the following condition
\begin{equation}
 \label{23}
\hat{J}_0 < R_m.
\end{equation}
Thus, in this case there is no phase transitions at all temperatures.
\end{theorem}
This statement is an adaptation of Theorem 7.3.1, page 346 in \cite{[A3]}.
Thus, (\ref{23}) is the condition under which the mentioned quantum stabilization holds.
In view of (\ref{V8}), it can be satisfied by picking small enough $m$.

To relate
(\ref{23}) with (\ref{S4}), we use the following result, see Theorem 7.1.1, page 338 in \cite{[A3]}.
\begin{theorem}
  \label{4tm}
Let $\nu$ and $V$ be as in Theorem \ref{3tm}. Then the rigidity parameter obeys the estimate
\begin{equation}
  \label{S5}
  R_m \leq \frac{1}{4 m \vartheta_*^2}.
\end{equation}
\end{theorem}
This estimate clearly shows that $R_m$ describes the tunneling motion between the wells. For if
$\vartheta_*$ is big, cf. (\ref{R2}), then $R_m$ gets small. On the other hand, for small $\vartheta_*$,
the bound (\ref{S5})  becomes inessential, and hence $R_m$  can be arbitrary. For such $\vartheta_*$,
we lose the control of $R_m$. This includes the case of convex $V$ where $\vartheta_*=0$, as it is for
harmonic crystals.

Thus, if (\ref{23}) holds, then
\begin{equation}
  \label{S6}
  Q < \varkappa:= 4 m \vartheta_*^2R_m \leq 1.
\end{equation}
It is known, cf. Proposition 6.3.5, page 308 in \cite{[A3]},
that $d \theta (d) >1$, and
\begin{equation}
 \label{S7}
 \frac{1}{d-1/2} < \theta (d) <  \frac{1}{d-1},
\end{equation}
for $d\geq 4$.
Therefore, for the phase transition to be suppressed at all temperatures, it suffices that (\ref{S6}) holds,
whereas (\ref{S4}) is the corresponding necessary condition. By (\ref{S7}), the right-hand side of (\ref{S4}) is
close to one for big $d$. Thus, for such $d$ the gap between these two conditions is close to $1 - \varkappa$.

\subsubsection{Normality of fluctuations in the $\nu >1$ case}

Unfortunately, we cannot extend Theorem \ref{3tm} to the vector case $\nu>1$. The main reason for this is that
the eigenvalues of the corresponding $H_x$ are no longer simple. The only result of this kind which so far we can get is
that the fluctuations of the displacements $q_x$ remain normal at all temperatures under a condition similar to (\ref{23}).

Here we again suppose that the potential $V$ is $O(\nu)$-symmetric. Then the operator which describes fluctuations
of the displacements of oscillators in a given $\Lambda$ is
\begin{equation}
  \label{S8}
F^{(\sigma)}_{j, \Lambda} = \frac{1}{|\Lambda|^{\sigma/2}} \sum_{x\in \Lambda} q_x^{(j)}, \quad j = 1, \dots , \nu, \quad \sigma \geq 1.
\end{equation}
Let us introduce the following parameter
\begin{equation*}
P_\Lambda^{(\sigma)} = \frac{1}{\beta} \sum_{j=1}^\nu \int_0^\beta \int_0^\beta \Gamma_{F^{(\sigma)}_{j, \Lambda} F^{(\sigma)}_{j, \Lambda}}(\tau, \tau') d \tau d\tau'.
\end{equation*}
Then $P_L =  P_{\Lambda_L}^{(2)}$, cf. (\ref{q20}). Clearly, if $P>0$, then $P_{\Lambda_L}^{(\sigma)} \to +\infty$ as $L\to +\infty$, for any
$\sigma \in [1, 2)$. We say that the fluctuations of the displacements of oscillators are {\it normal} if the Matsubara
functions (\ref{8z}) for the operators (\ref{S8}) with $\sigma =1$ remain bounded as $\Lambda \to \mathbb{Z}^d$.

Since $V$ is $O(\nu)$-symmetric, there exists $v(t)$, $t\geq 0$, such that $V(q) = v(|q|^2)$. Let $\Delta$ be the gap parameter (\ref{V5})
of the Hamiltonian
\begin{equation}
  \label{S10}
 \widetilde{H}_x := \frac{1}{2m} p_x^2 + v(q_x^2),
\end{equation}
of a one-dimensional oscillator with the same $m$ as the considered $\nu$-dimensional oscillator.
The following result is an adaptation of Theorem 7.3.5, page 350 in \cite{[A3]}.
\begin{theorem}
  \label{Stm}
Let $\nu>1$ be arbitrary, $V$ obey (\ref{3ba}), and let the condition (\ref{23}) be satisfied in which $\Delta$ is the gap parameter (\ref{V5})
of the Hamiltonian (\ref{S10}). Then the fluctuations of the displacements of oscillators in the corresponding model remain normal at all temperatures.
\end{theorem}
The proof of this result relies on  the so called scalar domination estimates obtained in Theorem 2.5.2, page 175 of \cite{[A3]}, see also \cite{[K2]}.

\subsubsection{Decay of correlations}

For simplicity, we suppose here that the model (\ref{1}) is of type (i), cf. (\ref{3}), with $\nu=1$. For this model,
the Matsubara function (\ref{q17}) describes the displacement-displacement correlations in the local periodic state (\ref{5a}). In the limit $L\to +\infty$, we obtain
periodic Gibbs states, possibly multiple. Under the stability condition (\ref{23}), there is only one such state. Then we set
\begin{equation}
  \label{S11}
 \Gamma_{xy} (\tau, \tau')= \lim_{L\to +\infty}\Gamma^L_{xy} (\tau, \tau').
\end{equation}
We know that $\Gamma_{xy} (\tau, \tau')$ should vanish in the limit $|x-y|\to \infty$, otherwise the corresponding state would be nonergodic.
The following statement, which is an adaptation
of Theorem 7.2.2, page 344 in \cite{[A3]}, describes the spatial decay of (\ref{S11}).
\begin{theorem}
  \label{S1tm}
Let the stability condition (\ref{23}) be satisfied. Then the correlation function (\ref{S11}) has the bound
\begin{widetext}
\begin{eqnarray}
  \label{S12}
 \Gamma_{xy} (\tau, \tau')
\leq \frac{1}{\beta (2 \pi)^d} \sum_{k\in \mathcal{K}} \int_{(- \pi, \pi]^d}
\frac{\exp\left[{\rm i}(p,x-y) + {\rm i} k(\tau - \tau') \right]}{m(\Delta^2 + k^2) + \Upsilon (p)} d p,
\end{eqnarray}
\end{widetext}
where
\begin{equation*}
 \Upsilon (p) =  2 \sum_{y} J_{xy} \left[ \sin(p,x-y)\right]^2,
\end{equation*}
and $\mathcal{K}$ is the set of all $k= (2\pi/\beta) \kappa$, $\kappa \in \mathbb{Z}$.
\end{theorem}
One observes that the right-hand side of (\ref{S12}) is the correlation function of a quantum harmonic crystal with the single-oscillator  rigidity $m\Delta^2$.
Thus, under the condition (\ref{23}) the decay of (\ref{S11}) is at least as strong as it is in the corresponding stable quantum harmonic crystal.
This result is valid for  arbitrary $J_{xy}$ obeying (\ref{J}). If the interaction has finite range, then
$ \Upsilon (p) \sim C|p|^2$ for some $C>0$. In this case, the spatial decay of the right-hand side of (\ref{S12}), and hence of
$ \Gamma_{xy} (\tau, \tau') $, is exponential, see Theorem 7.2.4, page 345 in \cite{[A3]}.

\section{Concluding remarks}

\subsection{To Section \ref{SS2}}
\subsubsection{The model}

In ionic crystals, the ions usually form massive complexes which
  determine the physical properties of the crystal, including the instability with respect to structural phase transitions.
  Such massive complexes can be considered as classical particles which obey the rules of classical statistical mechanics.
  At the same time, in a number of ionic crystals certain aspects of the phase transitions are apparently unusual from the point of view
  of classical physics, and can be explained only in a quantum-mechanical context. Such crystals contain light localized ions, and thus the mentioned features of their behavior point to the essential role of these light ions. In the simplest models of these
  crystals the motion of heavy complexes is ignored -- their only role is to create a potential field for the light particles.
  In this case, the relevant model parameters are those which describe this field, the masses of the light particles, and the interaction strength.
  Therefore, a consistent theory of phase transitions in such models should describe the influence of all these parameters.
We refer the reader to the survey article \cite{[S]} for more detailed arguments in favor of the model given in (\ref{1}).

The mathematical theory of a harmonic oscillator is quite standard, its updated and detailed presentation can be
found in \cite{[A3]}, pages 36--44. The theory of an anharmonic oscillator with a convex potential energy
is also quite standard. For instance, the case of
$H_x = H_x^{\rm har} + bq^{2r}$, $b>0$ and $r\geq 2$, was
carefully analyzed in \cite{[Ban]}. However, convex potentials do not have multiple wells, whereas the tunneling between the wells is the basic phenomenon in the studied substances. The theory outlined above in subsection
\ref{SS2.2}  among others employs quite recent results on Schr\"odinger operators obtained in \cite{[AB]}. By means of them we
found out for which potentials $V$ the domain of self-adjointness of $H_x$ can be
established exactly. This in turn allowed us to study the gap parameter (\ref{V5}) in great detail. Quite often,
see, e.g., \cite{[Anker]} and  \cite{[CDS]}, the description of the tunneling between the wells is concentrated on semiclassical expansions
for the gap parameter $E_1 - E_0$, which in our case would be in negative powers of $m= m_{\rm ph}/\hslash^2$.
Our analysis is essentially different, since it covers the case of `strong quantum effects' where semiclassical arguments are not applicable.

\subsubsection{Thermodynamic phases and the free energy}

The use of thermodynamic phases as mathematical objects is crucial in the
rigorous description of thermodynamic properties of a given model. In classical equilibrium
statistical physics, see \cite{[Ge]}, \cite{[Simon]}, and \cite{[Sinai]}, thermodynamic phases appear  as extreme elements of the set of Gibbs measures.
For quantum models described by bounded Hamiltonians, thermodynamic phases are pure KMS states, i.e., positive normalized linear functionals  on
algebras of quasi-local observables which obey the Kubo-Martin-Schwinger conditions, see Section 6.2 in \cite{[BR]}.
However, for quantum anharmonic crystals described by unbounded Hamiltonians, the direct construction of
such KMS states appears to be impossible, see, e.g., the discussion in \cite{[Ino]}.
In this situation, the construction of thermodynamic phases as path measures realized in \cite{[A3]}
seems to be the only possible way to settle the problem. It is performed by means of local Gibbs measures, cf.
(\ref{10}), within the general scheme developed in \cite{[Ge]}. The most significant point of our technique is
the representation of Matsubara functions in the form of path integrals, see (\ref{9}). The local Gibbs measure $\nu_\Lambda$ (\ref{10})
which appears in this representation is constructed from the corresponding Hamiltonian (\ref{3e}).
The harmonic part $\sum_{x\in \Lambda}H_x^{\rm har}$ of the latter is represented in (\ref{10}) by the measure $\chi_\Lambda$ (\ref{12}),
which is the Gibbs measure of noninteracting harmonic oscillators located in $\Lambda$. The interaction, the anharmonic part of $H_\Lambda$,
as well as the external field term are taken into account in the energy functional (\ref{11}).
By means of a technique developed in Chapter 2 of \cite{[A3]} the measure (\ref{10}) is approximated by
the Gibbs measure of a system of classical $\nu$-dimensional oscillators living on a $d+1$-dimensional lattice, see Section 2.1 {\it ibid}.
This extra dimension appears, in particular, when one approximates the integrals in (\ref{11})
by finite sums. Thus, the key feature of our method of studying thermodynamic states of the model (\ref{1})
is that we describe it as a system of classical oscillators performing `infinite dimensional oscillations'.

The most significant fact about the free energy density is its universality expressed in the equality (\ref{q9a}).
A deep fact behind it is the existence of van Hove sequences of subsets of $\mathbb{Z}^d$. It is known that such sequences do not exist
for non-amenable graphs, e.g., for a Bethe lattice. In this case, the phase diagram of an Ising model is completely different, compared to that
of this model on $\mathbb{Z}^d$. In particular, phase transitions can occur at nonzero $h$, cf. Theorem \ref{Ft} above and \cite{[Ly]}.
The mentioned universality, as well as the concavity of $F(h)$ which follows from the property (\ref{q12}), allow one
to get the polarization (\ref{q14}), and hence to study phase transitions.

\subsubsection{Phase transitions, order parameters, and infrared bounds}

When one deals with thermodynamic phases, the most general definition of a phase transition
is the multiplicity of such phases existing at the same values of the model parameters and
of the temperature and an external field, see Chapter 7 in \cite{[Ge]}. Another possibility
to define a phase transition, which goes back to L. D. Landau, is to use the differentiability property of
the free energy density $F(h)$ as a function of the external field $h$. One more possibility
is to employ an order parameter, cf. (\ref{q20}) and (\ref{q21}). For an Ising model,
all the three definitions are equivalent, see \cite{[Simon]}. In our case, however, we managed to establish only the fact stated in Theorem \ref{1tm}. For versions of the model (\ref{1}), phase transitions were previously discussed
mostly by means of the order parameter, see \cite{[MPZ],[SBS],[STZ],[VZ1],[VZ]}. The reason for this is that, in such a  case, one needs to control the
Matsubara functions (\ref{q17a}) only, which can be done not only by path measures.

The way of establishing phase transitions in quantum lattice models by means of infrared estimates was
suggested in \cite{[DLS]}. To the model (\ref{1}), in the present form it was developed  in \cite{[KKK]}. The key idea of
this method is to show that the spatial correlations do not decay to zero at large distances. This would point
to the existence of a Gibbs state which is not ergodic with respect to the lattice translations, and hence to the multiplicity
of thermodynamic phases. The key point here is the estimate in (\ref{q27}), where the first summand in the lower bound of
$D^L_{xy}$ is independent of $x$. Then the property in question is obtained if (\ref{q29}) is satisfied.

\subsection{To Section \ref{SS3}:}

\subsubsection{Phase transitions}

As was mentioned above, in our approach the system of quantum anharmonic oscillators is described as a system of classical
oscillators performing `infinite dimensional oscillations'. This allows us to apply here the original version of the
infrared estimates \cite{[FSS]} adapted to the infinite dimensional case.
The most transparent proof of the existence of a phase transition is in the $|q|^4$ case. The way of passing from (\ref{R3}) to
the key estimate (\ref{R4}) was suggested in \cite{[DLS]} where the function $f$ as in (\ref{18}) was introduced.
The condition (\ref{17}) is surprisingly simple. In fact, it points to a quantum phase transition which would occur at zero temperature.
If (\ref{17}) holds, then there exist multiple `ground states' of the model, whereas (\ref{S6})
implies that the `ground state' is unique. Of course, such a `ground state' ought to be defined as a mathematical object. For a simpler version of the model (\ref{1}), this was done in \cite{[AKMS]}. The extension of Theorem \ref{2tm} to more general models made in Theorem \ref{Rtm}
was done by means of a comparison method elaborated in \cite{[KP]} and based on a tedious analysis of path measures. The result of Theorem
\ref{R2tm} was first established in \cite{[KKK]}. Its main conclusion is that a first order  phase transition can occur without symmetry breaking. The way of getting this result was crucially based on the properties of path measures.

\subsubsection{Quantum effects}

As was mentioned above, the first paper where quantum effects in models like (\ref{1}) were discussed is \cite{[SBS]}.
However, only the systematic use of path measures, which allows for dealing with thermodynamic states as mathematical objects,
leads to the most complete description of such effects. Note also that a tedious analysis of the spectral properties
of a single anharmonic oscillator was crucial. The first paper where the notion of quantum stabilization was introduced is \cite{[A1]}.
The key parameter is the quantum rigidity (\ref{V7}). When it is large, the oscillator `forgets' about the details of the potential
energy in $H_x$ in the vicinity of the origin (including instability) and oscillates as if its equilibrium were stable, as in  the harmonic case.
The results of Theorem \ref{Stm} were obtained in \cite{[K1]} by means of the so called scalar domination inequalities obtained in \cite{[K2]}.
The result of Theorem \ref{S1tm} was obtained in \cite{[KK]}. The first systematic study of phase transitions and quantum effects in the model (\ref{1})
based on path integral methods was done in \cite{[KKK]}.

\section*{Acknowledgments}
The research presented in this paper was supported by the DFG through the SFB 701 ``Spektrale Strukturen und Topologische Methoden
in der Mathematik" and through the research project 436 POL 113/125/0-1.


%
%

%



\bibliographystyle{apsrmp}

\end{document}